\documentclass[twocolumn,aps,floats,letterpaper,floatfix,groupedaddress]{revtex4-2}
\usepackage{float}
\usepackage{afterpage}
\usepackage{amsmath}
\usepackage{amsfonts}
\usepackage{dsfont} 
\usepackage{amssymb}
\usepackage{makeidx}
\usepackage{graphicx}
\usepackage{subfig}
\usepackage{caption}
\usepackage{pgfplots}
\usepackage{array}
\usepackage{physics}
\usepackage{color}
\usepackage{mathptmx}
\usepackage{changes,enumitem,todonotes}
\usepackage{soul}
\usepackage[mathscr]{euscript}
\usepackage{natbib}
\usepackage[colorlinks=true,linkcolor=blue,urlcolor=blue,citecolor=blue]{hyperref}
\usepackage{tikz}
\usepackage[normalem]{ulem}
\usepackage[utf8]{inputenc}
\usepackage{hyperref}
\hypersetup{colorlinks=true,linkcolor=blue,urlcolor=blue,citecolor=blue} 
\usepackage{mathtools} 
\usepackage{graphicx}
\usepackage{lipsum} 
\usepackage{caption} 

\begin{document}

\title{Detecting winding and Chern numbers in topological matter using spectral function}
\author{Kiran Babasaheb Estake and Dibyendu Roy}
\affiliation{ Raman Research Institute, Bengaluru 560080, India}

\begin{abstract}
 We propose a method to directly probe bulk topological quantum numbers in topological matter by measuring the momentum-space single-particle spectral function (SPSF). Angle-resolved photoemission spectroscopy (ARPES) can detect SPSF and is often used to determine the bulk band structure of quantum materials. Here, we show that while one part of the momentum-space SPSF gives band structure, it also contains the knowledge of winding and Chern numbers of various topological materials. For this, we derive SPSF in different models of topological systems, such as the Kitaev model of topological superconductors, the long-range Su-Schrieffer-Heeger model, the Qi-Wu-Zhang model, the Haldane model on a hexagonal lattice and a four-band model that is a physical realization of the Kitaev chain and explain how to extract the winding or Chern numbers in different topological phases from the SPSF. Such information from SPSF seems experimentally accessible due to the recent advancement of ARPES and scanning tunneling spectroscopy techniques. While the detection of bulk topological quantum numbers using the SPSF works well for one-dimensional systems, it has certain limitations for higher-dimensional systems, where it must be complementary with another measurement to ensure the matter under investigation is topological, i.e., having a non-zero Chern number. 
\end{abstract}
\maketitle

\section{Introduction}
Topological concepts are often applied extensively in different scientific branches, particularly mathematics and physics. These concepts deal with the properties of a geometric object or manifold that remain invariant under continuous deformations. Such topological properties are essential in characterizing many physical phenomena and materials. There has been a surge of research activities in condensed matter physics since recognizing the quantum Hall effect \cite{thouless1998topological, QH} in solid-state systems as a topological phenomenon by \textcite{TKNN}. The topology of band structure in different phases of topological materials, such as insulators and superconductors, is identified through a bulk invariant, which distinguishes these different phases. For example, the winding number is a bulk topological invariant for one-dimensional (1D) topological systems. It is an integer denoting the total number of times a closed curve travels counterclockwise around a given point in a plane. The closed curve describes the endpoint of the Bloch vector of the band structure as crystal momentum runs through the Brillouin zone (BZ). Similarly, the Chern number, a bulk topological invariant for two-dimensional (2D) systems, denotes the number of times a closed 2D surface encloses a given point in three-dimensional space \cite{Asboth}.          

The winding number, Chern number, and other bulk topological invariants are shown to be related to many physical properties, such as polarization \cite{KingSmith1993} and electrical conductance \cite{Thouless1983}, in the topological phases of matter. Therefore, it is possible to find these bulk topological invariants by measuring the related physical properties, e.g., the measurement of quantized Hall conductivity in integer multiples of $e^2/h$ in a 2D electron system infers the topological quantum (Chern) number of the system \cite{thouless1998topological, QH}. Further, gapless edge modes can exist at the boundaries of topological matter, and the number of such edge modes is directly related to bulk topological invariants through the classic bulk-boundary correspondence \cite{Asboth}. Thus, local scattering measurements sensing such edge modes can also predict the topological invariants. Such an idea is employed in detecting the Zak phase of 1D topological insulators by measuring the reflection phase of scattered light \cite{Chan2014,Chan2015}. The detection of Majorana bound states and topological phase of 1D superconductors are also proposed following this idea \cite{Kitaev, BolechPRL2007, LawPRL2009, RoyPRB2012}.  

Here, we propose a method to directly detect the winding and Chern numbers in 1D and 2D topological systems. Our primary motivation for this study is to develop a unified description to probe the winding and Chern numbers in various topological systems, including topological insulators and superconductors. It is now clear that the detection of a zero-bias conductance peak does not ensure an unambiguous signature for the emergence of unpaired Majorana fermions at the edge of 1D topological superconductors, as such peaks can also appear due to various other physical mechanisms apart from topological superconductivity \cite{Kells2012,Roy2013}. Further, the fractional Josephson effect measurements in junctions of topological superconductors still need to be revised and clarified \cite{Bondyopadhaya2019}. Finally, our proposed method is expected to work for different topological insulators and superconductors as shown in Sec.~\ref{SPSF} and appendices. 

Here, we demonstrate that the high-quality data of single-particle spectral function (SPSF) can be used as a direct probe for the bulk topological quantum numbers of 1D and 2D quantum materials. The SPSF is experimentally accessible by angle-resolved photoemission spectroscopy (ARPES) \cite{ARPES1, ARPES2, Sobota2021} and by scanning tunneling spectroscopy (STS) \cite{Buchs2009, InteractingSSH}. The ARPES is routinely applied to determine the bulk band structure (energy-momentum dispersion) of quasiparticles and the many-body effects, such as Coulomb and spin interactions of electrons with other electrons and phonons, by detecting the SPSF \cite{Xia2009, Chen2010,SuXu2015}. The STS can probe the local density of states of a sample surface related to local SPSF in real space. It is a standard technique to detect the edge states of topological materials \cite{Yin21, Roushan2009, Alpichshev2010, Reis2017}. It can, in principle, also provide information in momentum-space spectral function via Fourier transformation from high-quality spatially-resolved measurements \cite{Buchs2009, Schneider2021}. We show below that while one part of the momentum-space SPSF gives band structure, it also contains the knowledge of winding and Chern numbers of topological materials. Such information from SPSF seems experimentally accessible due to the recent advancement of ARPES and STS techniques. 

While the detection of bulk topological quantum numbers using the SPSF works well for 1D systems, it has certain limitations for 2D systems. The SPSF can sometimes produce false positive or ambiguous results, which we discuss in Sec.~\ref{QWZ}. Because of these limitations, the SPSF-based detection must complement another measurement to ensure the matter under investigation is topological, i.e., having a non-zero Chern number. It can be done by probing local SPSF in real space by STS or ARPES to detect the edge states of topological materials. We also suggest another complementary method using the first-order coherence in Sec.~\ref{QWZ}. 

We organize the rest of the paper as follows. In Sec.~\ref{WNCN}, we introduce the basic concepts of the winding and Chern numbers for 1D and 2D quantum materials with or without chiral symmetry. We also briefly discuss SPSF here. In Sec.~\ref{SPSF}, we explicitly demonstrate how to find the winding and Chern number from the SPSF in three different models of topological systems. These models are the Kitaev model of 1D topological superconductors, the long-range Su-Schrieffer-Heeger (SSH) chain without chiral/sublattice symmetry, and the Qi-Wu-Zhang (QWZ) model. We conclude with a summary and outlook in Sec.~\ref{sum}. We discuss the trivial and non-trivial topology of 2D topological insulators in App.~\ref{TNT2Dtheta}. The details of the calculation of SPSF for the Kitaev chain and the QWZ model are given in App.~\ref{AppendixB}  and \ref{AppendixC}, respectively. App.~\ref{Disorder} discusses the SPSF in a disordered Kitaev chain. In App.~\ref{Haldane}, we study the SPSF for the Haldane model on the hexagonal lattice, and we discuss a four-band model that is the physical realization of the Kitaev chain in App.~\ref{Fourband}.

\section{Winding and Chern number from spectral function}\label{WNCN}
Here, we describe our scheme to find the winding and Chern number by measuring the SPSF for some generic Hamiltonian $H$ of 1D and 2D systems with two energy bands. The bulk momentum-space Hamiltonian $H({\bf k})$ can be written as
\begin{align}\label{g1}
  H=\sum_{{\bf k}}\Psi_{\bf k}^\dagger H({\bf k})\Psi_{\bf k},~~H({\bf k})=\vec{d}({\bf k})\cdot\vec{\sigma}.
\end{align} 
Here, \( H(\mathbf{k}) \) and \( \Psi_{\mathbf{k}} \) can describe both insulators and superconductors. For insulators, $H(\textbf{k})$ is the Bloch Hamiltonian, $\Psi_{\mathbf{k}}$ and $\Psi_{\mathbf{k}}^\dagger$ are annihilation operators and their Hermitian adjoint for momentum ${\bf k}$. For superconductors, $H(\textbf{k})$ describes the BdG Hamiltonian and  $\Psi_{\mathbf{k}}$ and $\Psi_{\mathbf{k}}^\dagger$ are the Nambu spinor and its Hermitian adjoint. The vector $\vec{d}({\bf k})$ is the Bloch vector and $\vec{\sigma}$ is the Pauli vector. First, we consider 1D quantum system with chiral symmetry and momentum $k$. Any one (say, $\lambda_1$) of the three components of $\vec{d}(k)$ vanishes due to the chiral symmetry. The periodicity of $ H(k)$ ensures that the endpoint of $\vec{d}(k)$ traces out a closed loop on the plane of two non-zero components (say, $\lambda_2,\lambda_3$) of $\vec{d}(k)$ as $k$ traverses the BZ, $k=-\pi \to \pi$. Here, $\lambda_1 \ne \lambda_2 \ne \lambda_3 \in (x,y,z)$. Further, the loop avoids the origin of the plane to host a bulk bandgap (pairing gap) for describing insulators (superconductors).   $H(k)$ provides a mapping from the BZ ($S^1$) to the $d_{\lambda_2}(k)$-$d_{\lambda_3}(k)$ plane. Considering only gapped Hamiltonians, we can normalize the vector $\vec{d}(k)$. This projects the closed loop on the unit circle centered at the origin. Hence, $H(k): S^1\rightarrow S^1$. The first homotopy group of $S^1$ is $\mathbb{Z}$. So, the topological phases of these systems are labelled by integer winding numbers. 

The winding number $\nu$ can be defined by unit vector $\hat{d}(k)=\vec{d}(k)/|\vec{d}(k)|$, which is the normalized Bloch vector \cite{Asboth}:
\begin{align}\label{g2}
  \nu=\frac{1}{2\pi}\int_{-\pi}^{\pi}\big(\hat{d}(k) \times \frac{d}{dk}\hat{d}(k)\big)_{\lambda_1}dk=\frac{1}{2\pi}\int_{-\pi}^{\pi} \frac{d\tilde{\theta}(k)}{dk}dk,
\end{align}
where, $\tilde{\theta}(k)$ is the angle between $\hat{d}(k)$ and any one non-zero component (say, $\lambda_2$) of $\hat{d}(k)$. The quantity $\tilde{\theta}(k)$ is of special importance for us, which would be clear once we lay out the results. 
 First, let us understand the notion of $\tilde{\theta}(k)$ geometrically. The vector $\vec{d}(k)$ traverses a closed loop in $d_{\lambda_2}(k)$-$d_{\lambda_3}(k)$ plane as $k$ traverses the BZ. When the winding number is $0$, $\vec{d}(k)$ does not wind around the origin and $\tilde{\theta}(k)$ varies within a proper subinterval of $[-\pi, \pi]$, e.g., if the closed loop lies in the third and fourth quadrant, then $\tilde{\theta}(k)$ will vary in $[-\pi+\tilde{\theta}_0(k),-\tilde{\theta}_0(k)]$, where $\tilde{\theta}_0(k)<\pi/2$. When the winding number is $1$, $\vec{d}(k)$ winds around the origin once, and $\tilde{\theta}(k)$ traverses the full interval $[-\pi, \pi]$ once. So, $\tilde{\theta}(k)$ contains the topology information or the winding number information. 

For 2D systems, we denote the momentum vector ${\bf k}=(k_x,k_y)$. The energies of the two bands are given by $E_{\pm}(k_x,k_y)=\pm|\vec{d}(\textbf{k})|$. The band touching occurs when $|\vec{d}(\textbf{k})|=0$, which sets the origin of the space spanned by the three components of $\vec{d}(\textbf{k})$. Again, $\vec{d}(\textbf{k})$ for a gapped $H({\bf k})$ provides a mapping from the BZ, which is a torus, to the space $\mathbb{R}^3-\{0\}$. The tip of  $\vec{d}(\textbf{k})$ also traces a deformed torus in the space $\mathbb{R}^3-\{0\}$ as $k_x$ and $k_y$ vary in the BZ (see App. \ref{TNT2Dtheta}). We find different Chern numbers depending on how many times and how this deformed torus encloses the origin \cite{Asboth}. For a gapped $H(\textbf{k})$, we can calculate the Chern number for each band of the 2D system using,
\begin{align}
Q^{\pm}=-\frac{1}{2\pi}\int_{BZ}dk_xdk_y\Bigl(\frac{\partial A_y^{\pm}}{\partial k_x}-\frac{\partial A_x^{\pm}}{\partial k_y}\Bigr),\label{ChernNo}\\
{\rm where}~~ A_{\chi}^{\pm}(\textbf{k})=i\bra{u_{\pm}(\textbf{k})}\partial_{k_{\chi}}\ket{u_{\pm}(\textbf{k})},
\end{align}
for $\chi=x,y.$ Here, $\ket{u_{\pm}(\textbf{k})}$ are the energy eigenstates of $H({\bf k})$ for two bands denoted by $\pm$. The concept of Chern number is directly related to the geometric (Berry) phase, where $A_{\chi}^{\pm}(\textbf{k})$ is the Berry connection, and the integrand in Eq.~\ref{ChernNo} is the Berry curvature. The Chern number is a gauge invariant quantity, and does not depend on the choice of gauge or unit cell of the system. Thus, the Chern number can be uniquely measured in each topological phase of matter. 
 
Let us consider a spherical coordinate system $(r, \Theta, \Phi)$ to denote the  tip of $\vec{d}(\textbf{k})$ in $d_x({\bf k})$-$d_y({\bf k})$-$d_z({\bf k})$ space, where the radial distance $r=|\vec{d}({\textbf k})|$, $\Theta(\textbf k)$ is the angle between $\vec{d}(\textbf k)$ and the $d_z(\textbf k)$ axis, and $\Phi(\textbf k)$ is the angle between the projection of $\vec{d}(\textbf k)$ on the $d_x({\textbf k})$-$d_y({\textbf k})$ plane and the $d_x({\bf k})$ axis. The accessible values of the polar angle $\Theta({\textbf k})$ can differentiate between the trivial and the non-trivial phase of 2D systems in consideration. When a system is in a non-trivial topological phase, the deformed torus in $d_x({\textbf k})$-$d_y({\textbf k})$-$d_z({\textbf k})$ space encloses the origin. Then there exists a region of the 2D BZ in which as $k_x(k_y)$ varies from $-\pi$ to $\pi$ for a fixed $k_y(k_x)$, $\Theta$ can access all values in interval $[0,\pi]$. On the other hand, the origin is outside of the deformed torus for the system in a trivial phase, and there is no such region in the 2D BZ where $\Theta$ behaves in the manner stated above for any fixed $k_y$ or $k_x$. Instead, $\Theta$ would traverse a proper subinterval of $[0,\pi]$ as $k_x$ (or $k_y$) varies from $-\pi$ to $\pi$ for any fixed $k_y$ (or $k_x$). See App. \ref{TNT2Dtheta} for more details. 

For 1D systems without chiral symmetry, $\vec{d}(k)$ is not restricted on a plane. Then, the winding number cannot be defined to characterize their topological features. One can instead introduce a continuously varying cyclic parameter, say $\varphi$. This is commonly done in topological insulators to increase or decrease the dimension of the system; it is called dimensional extension or reduction. In such a case, this extra parameter can be thought of as another momentum along with the original momentum ($k$) corresponding to the physical dimension. The Chern number can be calculated for such 1D systems by substituting $\varphi$ for $k_y$ and $k$ for $k_x$ in Eq.~\ref{ChernNo} to quantify their bulk topology. We shall use this description in the long-range SSH chain without chiral symmetry and its SPSF below.

We now discuss extracting knowledge of $\nu$ and $Q^{\pm}$ from the SPSF. The correlation functions or Green's functions, which are time-ordered correlators of two field operators, are vital tools to evaluate crucial physical observables of any system. Here, we study the retarded Green's function and the associated spectral function for topological matters. The retarded Green's function for systems at zero temperature is defined as \cite{Bruus2004}
\begin{align}\label{retG}
G^R_{\alpha \beta }(\textbf{r}_1,t_1;\textbf{r}_2,t_2)=-i\theta(t_1-t_2)\bra{\mathcal{G}}\{c_{\textbf{r}_1,\alpha } (t_1), c^\dagger_{\textbf{r}_2,\beta }(t_2)\}\ket{\mathcal{G}},
\end{align}
where, $\theta(t_1-t_2)$ is the Heaviside step function, and $\ket{\mathcal{G}}$ is the half-filled many-particle ground state.  $c^\dagger_{\textbf{r}_2,\beta }~(c_{\textbf{r}_2,\beta })$ is a fermionic creation (annihilation) operator at sublattice $\beta$ of unit cell at position $\textbf{r}_2 \equiv ({r_2}_{x},{r_2}_{y})$, and $\{c_{\textbf{r}_1,\alpha } (t_1), c^\dagger_{\textbf{r}_2,\beta }(t_2)\}$ denotes the anti-commutator of $c_{\textbf{r}_1,\alpha } (t_1)$ and $c^\dagger_{\textbf{r}_2,\beta }(t_2)$, where $c_{\textbf{r}_1,\alpha } (t_1)=e^{iHt_1}c_{\textbf{r}_1,\alpha } (0)e^{-iHt_1}$.  We call this Green's function diagonal for $\alpha=\beta$, and off-diagonal when $\alpha\neq\beta$. The Green's function depends only on $(\textbf{r}_1-\textbf{r}_2)$ and $(t_1-t_2)$ for space and time translational invariant systems. Using this fact, we can take the Fourier transform of the retarded Green's function with respect to $(\textbf{r}_1-\textbf{r}_2)\equiv\textbf{r}$ and $(t_1-t_2)\equiv t$. Let us call it $G^R_{\alpha \beta}(\textbf{k},\omega)$. 

\begin{align}
G^R_{\alpha \beta}(\textbf{k},\omega)=\sum_{\textbf{r}} e^{i\textbf{k}\cdot\textbf{r}}\int_{-\infty}^\infty dt e^{i\omega t}G^R_{\alpha \beta}(\textbf{r}, t).
\end{align}
The momentum-space SPSF $A_{\alpha \beta}(\textbf{k},\omega)$ is given as \cite{Bruus2004} 
\begin{align}
A_{\alpha \beta}(\textbf{k},\omega)=-2 \text{Im} G^R_{\alpha \beta}(\textbf{k},\omega).
\end{align}
Thus, the SPSF is diagonal or off-diagonal, depending on the retarded Green's function. The diagonal SPSF is a gauge invariant quantity that can be experimentally probed. We show below that the diagonal SPSF provides the knowledge on topology for the long-range SSH chain and the QWZ model. Since there is no sublattice structure, the choice of diagonal or off-diagonal SPSF does not arise for the Kitaev model of topological superconductors.

\section{Spectral function of topological systems}\label{SPSF}
We now derive the SPSF for three different models of 1D and 2D topological insulators and superconductors. For this, we choose the Kitaev chain \cite{Kitaev} representing a $p$-wave topological superconductor, the long range SSH chain \cite{Generalized} and the 2D QWZ model \cite{QWZ2006, Asboth}. Each of these models has non-trivial topological phases with nonzero winding or Chern numbers. We discuss these non-trivial topologies along with the finding of winding and Chern number from the SPSF in this section.

\subsection{Kitaev chain}\label{Kitaev}
\textcite{Kitaev} proposed a simple 1D model of topological superconductors of spinless fermions with $p$-wave superconducting pairing \cite{Kitaev, AliceaReview2012}. The Kitaev Hamiltonian for $N$ sites reads 
\begin{align}\label{5}
	H_K=  -\sum_{j=1}^{N}\mu c_{j}^\dagger c_{j}+\sum_{j=1}^{N-1} \Bigl(\Delta c_{j} c_{j+1}-Jc_{j}^\dagger  c_{j+1}+h.c.\Bigr),
\end{align}
where, $c_{j}^\dagger$ and $c_j$ are, respectively, the creation and annihilation operators of fermion at site $j$. The parameters $\mu, J$ and $\Delta$ are, respectively, the onsite chemical potential, nearest-neighbour hopping, and the mean-field superconducting pairing. We choose all of them to be real positive numbers. The pairing term breaks  $U(1)$ symmetry or total fermion number conservation of $H_K$. Nevertheless, the parity (oddness or evenness) of the total number of fermions is conserved. We take periodic boundary condition (PBC) to write $H_K$ in momentum space by taking Fourier transforms of the creation and annihilation operators, e.g., $c_j=\frac{1}{\sqrt{N}}\sum_ke^{ikj}\tilde{c}_k$. In BdG form, $H_K$ can be written as $H_K=\sum_{k}\Psi_k^\dagger H_K(k)\Psi_{k}$, where $\Psi_k=\begin{pmatrix} \tilde{c}_{k}& \tilde{c}_{-k}^{\dagger}\end{pmatrix}^T$, and
\begin{align}\label{7}
H_K(k)=\begin{pmatrix}{-\mu/2-J\cos{k}} &-{i\Delta \sin{k}}\\{i\Delta \sin{k}}&{\mu/2+J\cos{k}}\end{pmatrix}=\vec{d}(k)\cdot\vec{\sigma}.
\end{align}
The eigenvalues of this Hamiltonian are given by $E_{\pm}(k)=\pm\sqrt{\Delta^2 \sin^2k+(\mu/2+J\cos{k})^2}=\pm E(k)$, where $E(k)=|\vec{d}(k)|=\sqrt{\Delta^2 \sin^2k+(\mu/2+J\cos{k})^2}$. So, there is a pairing gap $2E(k)$. This gap vanishes when $|E_{\pm}(k)|=0$, which occurs when $|\vec{d}(k)|=0$.

$H_K(k)$ satisfies chiral symmetry, particle-hole symmetry, and time reversal symmetry. The chiral symmetry operator in this case is $\Gamma=\sigma_x$, and the chiral symmetry is described as  $\Gamma H_K(k) \Gamma^{-1}=-H_K(k)$. The particle-hole symmetry is described as $\mathcal{C}H_K(k)\mathcal{C}^{-1}=-H_K(-k)$, with $\mathcal{C}=\sigma_x \mathcal{K}$, where $\mathcal{K}$ is the time reversal operator, specifically the complex conjugation operator. The time reversal symmetry is described as $\mathcal{K}H_K(k)\mathcal{K}^{-1}=H_K(-k)$. The components of the Bloch vector here are $d_x(k)=0$, $d_y(k)=\Delta \sin{k}$, $d_z(k)=-\mu/2-J\cos{k}$. There is no $d_x(k)$ component to $\vec{d}(k)$ as a consequence of chiral symmetry, and $\vec{d}(k)$ is restricted in $d_y(k)$-$d_z(k)$ plane. This makes Kitaev chain a symmetry-protected topological superconductor. $H_K(k)$ provides a mapping from the BZ to $d_y(k)$-$d_z(k)$ plane, and it forms an ellipse, which can be used to define a winding number according to \ref{g2}. The Kitaev chain supports $\nu=0$ (trivial phase) and $\nu=1$ (non-trivial phase) depending on the values of $\mu$ and $J$. For $\mu,J>0$,  the Kitaev chain is in trivial phase when $\mu>2J$, and it is in non-trivial phase supporting two zero-energy Majorana modes at the edges of the open chain, when $\mu<2J$.

As the crystal momentum $k$ traverses the BZ, the endpoint of $\vec{d}(k)$ traces out an ellipse in  $d_y(k)$-$d_z(k)$ plane, which encloses (excludes) the origin when $\mu<2J ~(\mu>2J)$. We now define the angle  between $\vec{d}(k)$ and $d_y(k)$ axis by $\tilde{\theta}(k)$. Thus, we find $\sin \tilde{\theta}(k)=-(\mu/2+J\cos{k})/E(k)$ and $\cos \tilde{\theta}(k)=\Delta \sin{k}/E(k)$. The angle $\tilde{\theta}(k)$ contains the information of topology of the Kitaev chain's band structure. For $\nu=0$, since the ellipse lies in the third and fourth quadrant, $\tilde{\theta}(k)$ traverses the interval $[-\pi+\tilde{\theta}_0, -\tilde{\theta}_0]$, where $\tilde{\theta}_0$ is some angle less than $\pi/2$. For $\nu=1$, $\tilde{\theta}(k)$ traverses the full interval of $[-\pi,\pi]$ once. 

We diagonalize $H_K(k)$, and write it in terms of the normal modes as
\begin{align}\label{dHK}
	  H_K=\sum_{k}\Bigl(E(k)\tilde{\eta}_{k}^\dagger \tilde{\eta}_{k} -E(-k)\tilde{\eta}_{-k}\tilde{\eta}_{-k}^\dagger \Bigr),
\end{align}
where, $\tilde{\eta}_{k}=X(k)\tilde{c}_{k}+Y^*(k)\tilde{c}_{-k}^\dagger$ are the Bogoliubov excitations. Here, $X(k)=\cos \tilde{\theta}(k)/\sqrt{2(1-\sin\tilde{\theta}(k))}$ and $Y(k)=i\sqrt{(1-\sin \tilde{\theta}(k))/2}$. We define the ground state $|\mathcal{G}\rangle$ as the vacuum of the Bogoliubov excitations that is annihilated by all $\tilde{\eta}_k$'s, e.g., $\tilde{\eta}_k \ket{\mathcal{G}}=0$.

The retarded Green's function for the Bogoliubov vacuum $\ket{\mathcal{G}}$ is given by,
\begin{align}\label{23}
	G^R(x_1,t;x_2,0)=-i\theta(t)\bra{\mathcal{G}}\{c_{x_1}(t),c_{x_2}^\dagger(0)\}\ket{\mathcal{G}},
\end{align}
which does not have any sublattice index for the Kitaev chain. From this point onward, we choose $t_2=0$, and denote $t_1=t$ in Eq. \ref{retG}. We find in App.~\ref{AppendixB}:
\begin{align}\label{24}
G^R(x_1,t;x_2,0)&=-i\frac{\theta(t)}{2N}\sum_{k\in BZ} e^{ik(x_1-x_2)}\Bigl[e^{-iE(k)t}\Bigl(1+\sin\tilde{\theta}(k)\Bigr)\notag\\
	&+e^{iE(k)t}\Bigl(1-\sin\tilde{\theta}(k)\Bigr)\Bigr].
\end{align}
Thus, the SPSF for the Bogoliubov vacuum is found to be 
\begin{align}\label{SPSFK}
	A(k,\omega)&=-2\text{Im}G^R(k,\omega)\notag\\
	&=\pi(1+\sin\tilde{\theta}(k))\delta(\omega-E(k))\notag\\
	&+\pi(1-\sin\tilde{\theta}(k))\delta(\omega+E(k)),
\end{align}
where, $G^R(k,\omega)$ is the Fourier transform of $G^R(x_1,t;x_2,0)$ with respect to $(x_1-x_2)$ and $t$ (see App. \ref{AppendixB}).
\begin{figure}[h]
  \centering
  \includegraphics[width=\linewidth]{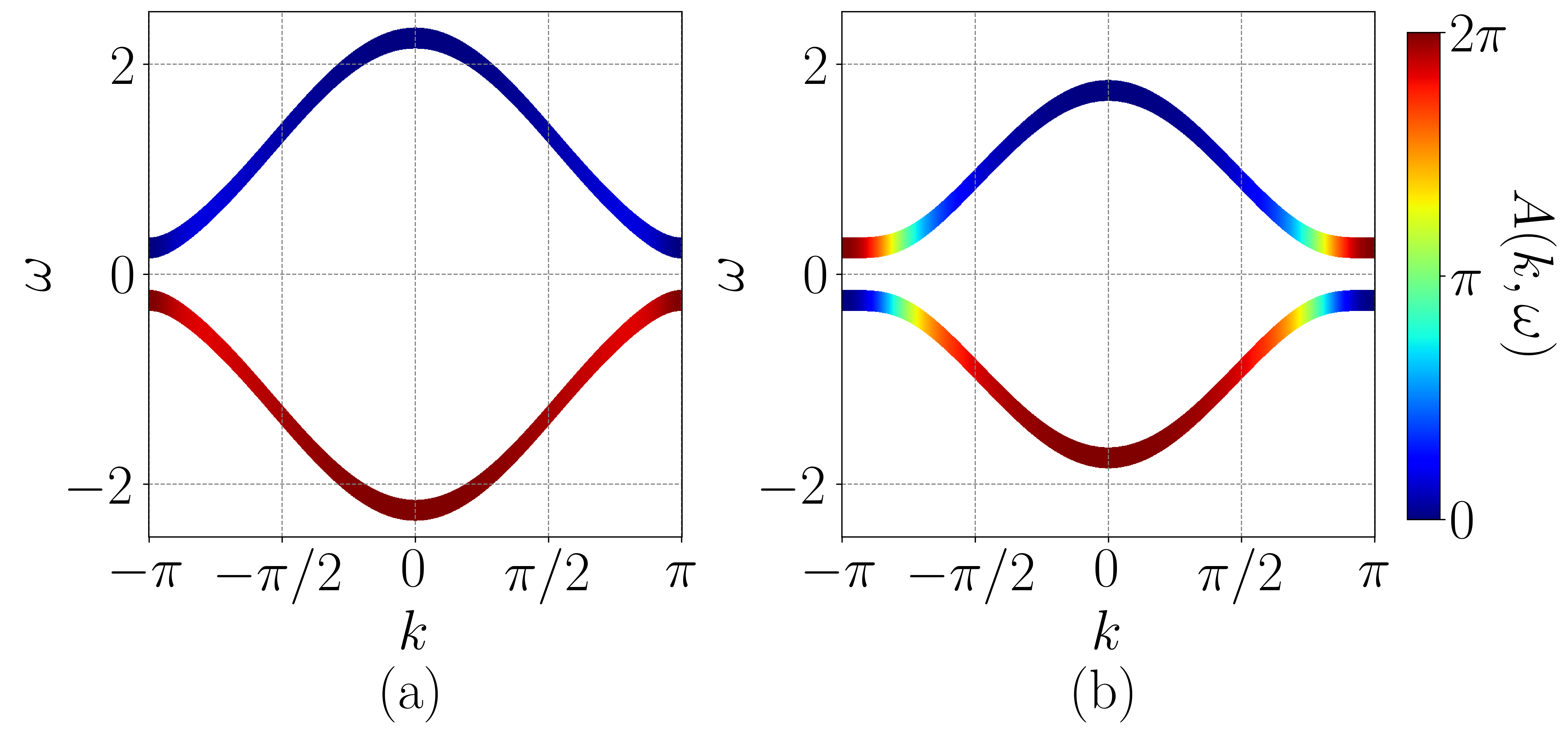}
  \caption{SPSF $A(k,\omega)$ with $k,\omega$ for the Kitaev chain in (a) trivial phase with $\mu=2.5, J=1, \Delta=0.5$, and (b) non-trivial phase with $\mu=1, J=1, \Delta=0.5$. For trivial phase, the value of SPSF is greater (less) than $\pi$ in the entire BZ for lower (upper) band.  For non-trivial phase, $\tilde{\theta}(k)$ winds around the origin once, which leads to $(1\pm\sin\tilde{\theta}(k))$ crossing $1$ twice; thus there are two $\pi$ crossings of SPSF for both the bands.}
  \label{fig.3}
\end{figure}
The emergence of $\tilde{\theta}(k)$ in the SPSF in Eq.~\ref{SPSFK} signals its sensitivity to different topological phases of the Kitaev chain. We can determine the winding number $\nu$ of each topological phase by detecting the momentum-space SPSF. We plot $A(k,\omega)$ in Fig.~\ref{fig.3} in the trivial and non-trivial phase of the Kitaev chain, which shows qualitative differences in color across the BZ in the two phases. The line shapes or contours of the plots give the energy-momentum dispersion ($\omega=\pm E(k)$) of the bands, where $\delta(\omega\pm E(k))$ functions are approximated by box functions to mimic the broadening of the SPSF due to electron's coupling to other degrees of freedom, e.g., phonon \cite{Bruus2004}. We now discuss how measuring such plots of spectral function experimentally can infer about the winding number by counting the number of ``$\pi$ crossings'' of $A(k,\omega)$. A $\pi$ crossing is when $A(k,\omega)$ crosses the value $\pi$ in any energy band. This occurs due to a change in sign of $\sin \tilde{\theta}(k)$. We show here that the number of such $\pi$ crossings in each band is double of $\nu$ of that topological phase. Fig.~\ref{fig.3}(a) for $\nu=0$ shows no such $\pi$ crossings, and $A(k,\omega)$ is always greater than $\pi$ for the lower band and less than $\pi$ for the upper band for all $k$. For $\nu=0$, since $\tilde{\theta}(k)$ lies in the third and fourth quadrant, $ \sin \tilde{\theta}(k)$ is negative for all $k$ in the BZ. Therefore, $(1-\sin\tilde{\theta}(k))$ is greater than $1$, and $(1+\sin\tilde{\theta}(k))$ is less than $1$ for all $k$. Thus, $A(k,\omega)$ is greater than $\pi$ for lower band and less than $\pi$ for upper band, for all $k$ in the BZ. For $\nu=1$, the value of $A(k,\omega)$ crosses $\pi$ twice in each band. This can be understood as a consequence of topology alone, in terms of $(1\pm\sin\tilde{\theta}(k))$ factors. We saw that $\tilde{\theta}(k)$ traverses the interval $[-\pi,\pi]$ once for $\nu=1$. Thus, $ \sin\tilde{\theta}(k)$ would become zero twice, at $\tilde{\theta}(k)=0$ and at $\tilde{\theta}(k)=\pi$. As a result of this, the factors $(1\pm\sin\tilde{\theta}(k))$ become $1$ twice, giving rise to two $\pi$ crossings in Fig.~\ref{fig.3}(b), for both the bands.

\subsection{Long-range SSH chain}\label{ExtendedSSH}
\begin{figure}[h]
  \centering
    \includegraphics[width=\linewidth]{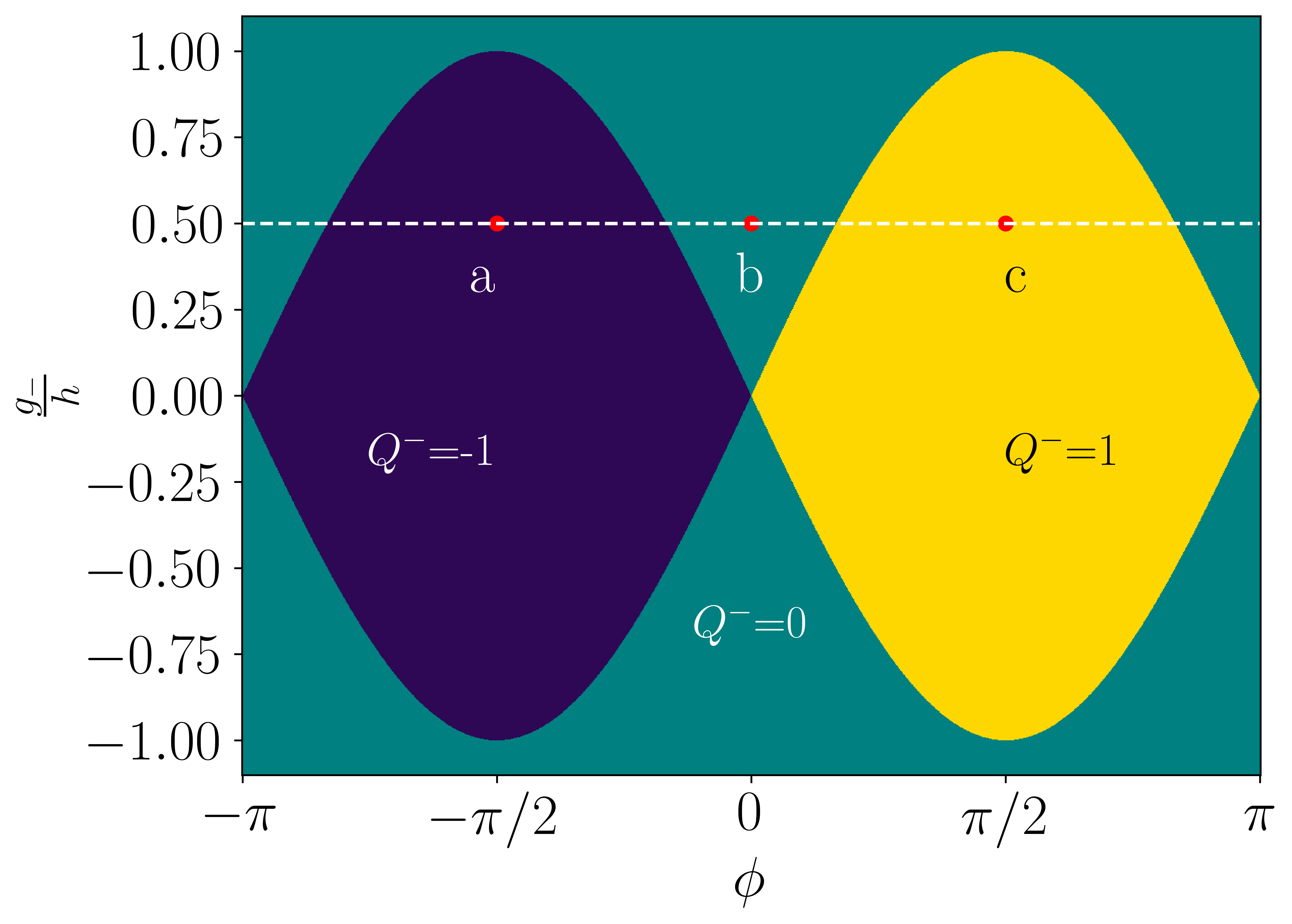}
    \caption{A cartoon phase diagram of the long-range SSH model showing three topological phases with Chern number $Q^{-} = 0, \pm 1$. The  red dots denote parameters for three different topological phases for which we discuss the features of SPSF.}\label{ESSH_phasediagram}
\end{figure}

The SSH model \cite{SSH, Asboth, TopolectricalCircuits} was originally proposed as a model for poly-acetylene chain, in which only hoppings among different sublattices are considered, which makes it a sublattice symmetric or chiral symmetric model. Nevertheless, we can also add hoppings among the same sublattice sites to the SSH chain to break chiral symmetry \cite{Generalized}. Such a generalization of SSH chain leads to an effective 2D model, if the hoppings are appropriately parametrized by a cyclic parameter. It supports topological phases with non-zero Chern numbers, and has a phase diagram similar to the Haldane model \cite{Haldane, Generalized}. The minimal Hamiltonian of this long-range SSH chain of size $N$ is given by, 
\begin{align}
H_{L}=&\sum_{j=1}^N\bigl(J_0 c_{j,\alpha}^\dagger c_{j,\beta}+h.c.\bigr)+\sum_{j=1}^{N-1}\bigl(J_1c_{j+1,\alpha}^\dagger c_{j,\beta}+h.c.\bigr)\notag\\
&+\sum_{j=1}^{N-1}\bigl(J_{\alpha} c_{j,\alpha}^\dagger c_{j+1,\alpha}+J_{\beta}c_{j,\beta}^\dagger c_{j+1,\beta} +h.c. \bigr),
\end{align}
where, $c_{j,\alpha}^\dagger$ and $c_{j,\alpha}$ are the creation and annihilation operators of fermion at site $j$ and sublattice $\alpha$. We parametrized $J_0=1+\delta \cos\varphi$ and $J_1=1-\delta \cos\varphi$. Here, $\delta$ controls the dimerization, and $\varphi$ is a cyclic parameter which changes continuously from $-\pi$ to $\pi$. For PBC, we get the momentum-space Hamiltonian by introducing  $c_{j,\alpha}=(1/\sqrt{N})\sum_ke^{ikj}\tilde{c}_{k,\alpha}$, 
\begin{align}
H_{L}=\sum_k\Psi_k^\dagger H_L(k,\varphi) \Psi_k,~H_L(k,\varphi)=\begin{pmatrix}2J_{\alpha}\cos k&J_0+J_{1}e^{-ik}\\J_0+J_{1}e^{ik}&2J_{\beta}\cos k\end{pmatrix},
\end{align}
where, $\Psi_k^\dagger=\begin{pmatrix} \tilde{c}^\dagger_{k,\alpha}& \tilde{c}^\dagger_{k,\beta}\end{pmatrix}$. 
We further parametrize the long-range hoppings as $J_{\alpha}=g_{\alpha}+h\cos(\varphi+\phi)$ and $J_{\beta}=g_{\beta}+h\cos(\varphi-\phi)$. $H_{L}(k,\varphi)$ can be written as 
\begin{align}
H_{L}(k,\varphi)=d_0(k,\varphi)I+\vec{d}(k,\varphi)\cdot\vec{\sigma},
\end{align}
where, $d_0(k,\varphi)=2\cos k(g_++h\cos\varphi\cos\phi)$, $d_x(k,\varphi)=(1+\delta\cos\varphi)+(1-\delta\cos\varphi)\cos k$, $d_y(k,\varphi)=(1-\delta\cos\varphi)\sin k$ and $d_z(k,\varphi)=2\cos k(g_--h\sin\varphi\sin\phi)$, with $g_{\pm}=(g_{\alpha} \pm g_{\beta})/2$.
The energies of the upper and lower bands are: $E_\pm(k,\varphi)=2\cos k(g_++h\cos\varphi\cos\phi)\pm \Delta E(k,\varphi)$, where $(\Delta E(k,\varphi))^2=|\vec{d}(k,\varphi)|^2=4\cos^2k(g_--h\sin\varphi\sin\phi)^2+2(1+\cos k)+2\delta^2\cos^2\varphi(1-\cos k)$.  The gap between the bands is $2\Delta E(k,\varphi)$. The gap closing ($\Delta E=0$) occurs when $g_-=\pm h\sin\phi$, for $\varphi=\pm \pi/2$ and $k= \pi$. We can calculate the Chern number for this model by treating $\varphi$ as another cyclic parameter in addition to $k$. This model has three different topological phases characterized by Chern number $Q^{-}=0,\pm1$ for the lower band. 
The topological phase diagram of this model showing these three phases is given in Fig.~\ref{ESSH_phasediagram}, which shows the phase boundaries denoted by $g_-/h=\pm \sin\phi$ \cite{Generalized}. 

\begin{figure*}[t]
  \centering
  {\includegraphics[width=1\textwidth]{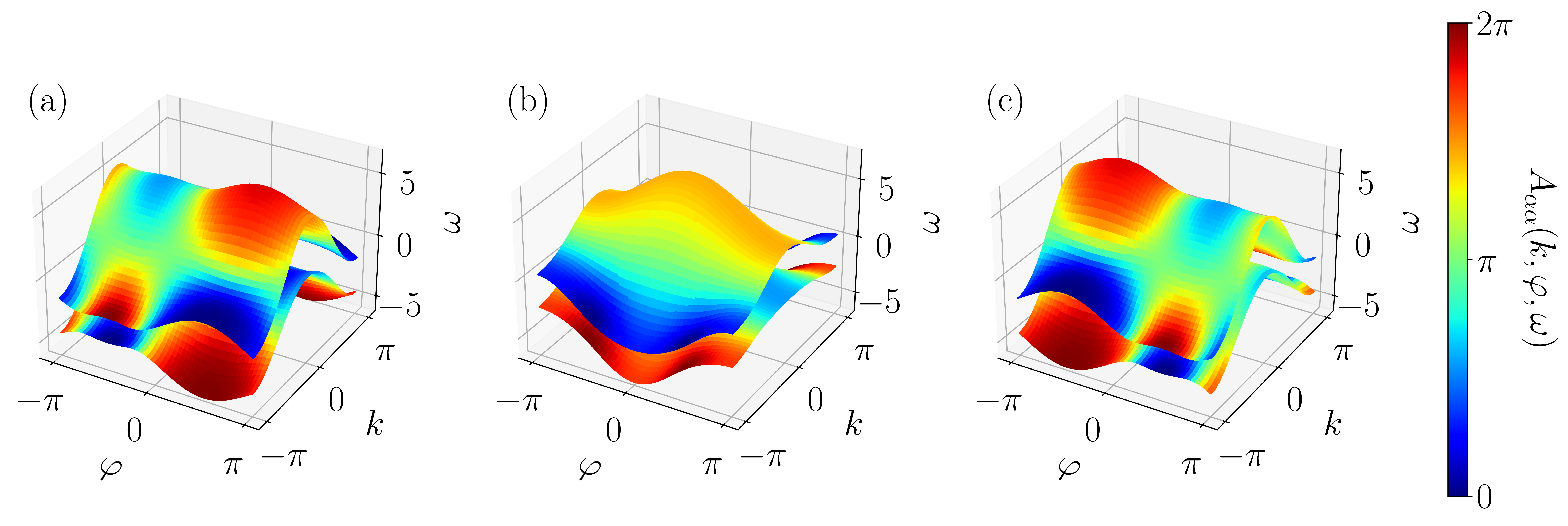}}
  \captionsetup{justification=justified,singlelinecheck=false}
  \caption{Diagonal SPSF $(A_{\alpha \alpha}(k,\varphi,\omega))$ with $k$, $\varphi$, and $\omega$ for Chern number $Q^{-}=0,\pm 1$ of the long-range SSH chain. We set $g_{\alpha}=2$, $g_{\beta}=1$, $h=1$, $\delta=0.5$. (a) $Q^{-}=-1$ (dot a of Fig.~\ref{ESSH_phasediagram}) at $\phi=-0.5 \pi$. At $k=\pi$, SPSF takes values $0\rightarrow 2\pi \rightarrow 0$ for both upper and lower bands, as $\varphi$ varies in 2D BZ, (b) $Q^{-}=0$ (dot b of Fig.~\ref{ESSH_phasediagram}) at $\phi=0$. SPSF is either greater than or less than $\pi$ for any fixed $k$, for either energy bands. (c) SPSF for $Q^{-}=1$ (dot c of Fig.~\ref{ESSH_phasediagram}) at $\phi=0.5 \pi$. SPSF, similar to the one in (a), varies as $0\rightarrow 2\pi\rightarrow 0$ at $k=\pi$ in 2D BZ.}\label{SPSFE}
\end{figure*}

The Hamiltonian $H_{L}(k,\varphi)$ allows a map from an effective 2D BZ (torus) to the $d_x$-$d_y$-$d_z$ space. As $k$ and $\varphi$ vary from $-\pi$ to $\pi$, the tip of $\vec{d}(k,\varphi)$ traces out a deformed torus in this space (see App. \ref{TNT2Dtheta}). When the long-range SSH chain is in a non-trivial (trivial) phase, the deformed torus encloses (excludes) the origin $\{0,0,0\}$. Let us consider the spherical coordinate system $(r, \Theta, \Phi)$ in $d_x$-$d_y$-$d_z$ space. The accessible values of  $\Theta(k,\varphi)$ in the topologically trivial and non-trivial phases are described in Sec.~\ref{WNCN} and App. \ref{TNT2Dtheta}. We find $\Theta(k,\varphi)$ explicitly appearing in the SPSF of this model. Thus, we can identify different topological phases of this model, based on the accessible values of $\Theta(k,\varphi)$, from the SPSF. 

We find that the diagonal SPSF calculated for the same sublattice (say $\alpha$) has the signature of topology in this model. Thus, we evaluate the following zero temperature diagonal retarded Green's function in the half-filled many particle ground state $\ket{\mathcal{G}}$: 
\begin{align}
&G^R_{\alpha\alpha}(x_1,t;x_2,0)\notag\\
&=-i\theta(t)\bra{\mathcal{G}}\{c_{x_1,\alpha}(t),c_{x_2,\alpha}^\dagger(0)\}\ket{\mathcal{G}}\notag\\
&=-i\frac{\theta(t)}{2N}\sum_{k\in BZ}e^{ik(x_1-x_2)}\Bigl[e^{-iE_+(k,\varphi)t}\Bigl(1+\cos\Theta(k,\varphi)\Bigr)\notag\\
&~~~+e^{-iE_-(k,\varphi)t}\Bigl(1-\cos\Theta(k,\varphi)\Bigr)\Bigr],
\end{align}
where, $\cos \Theta(k,\varphi)=d_z(k,\varphi)/|\vec{d}(k,\varphi)|$.
The diagonal SPSF at zero temperature depends on $\varphi$ too. 
\begin{align}
	A_{\alpha \alpha}(k,\varphi,\omega)&=-2\text{Im}G_{\alpha \alpha}^R(k,\varphi,\omega)\notag\\
  &=\pi\Bigl(1+\cos\Theta(k, \varphi)\Bigr)\delta(\omega-E_+(k,\varphi))\notag\\
  &~~~+\pi\Bigl(1-\cos\Theta(k, \varphi)\Bigr)\delta(\omega-E_-(k,\varphi)),\label{SPSF_LR}
\end{align}
where, $G_{\alpha \alpha}^R(k,\varphi,\omega)$ is the Fourier transform of $G^R_{\alpha\alpha}(x_1,t;x_2,0)$ with respect to $(x_1-x_2)$ and $t$.

Because of the presence of the factors $1\pm\cos\Theta(k,\varphi)$ in the SPSF, we expect it to show qualitatively different behaviour in trivial and non-trivial phase. In Fig.~\ref{SPSFE}, we plot the diagonal SPSF in Eq.~\ref{SPSF_LR} as a function of $k,\varphi$ and $\omega$ for three different sets of parameters (three dots in Fig.~\ref{ESSH_phasediagram}) representing three different topological phases. The delta functions in Eq.~\ref{SPSF_LR} give the band structure of the long-range SSH chain, and the strengths of these delta functions are represented by the color plots in Fig.~\ref{SPSFE}. These plots of SPSF are qualitatively different for zero and non-zero Chern numbers. Figs.~\ref{SPSFE}(a,c) show the SPSF for a non-trivial phase. For these cases,  $\Theta(k,\phi)$ takes all values in $[0,\pi]$ as $\varphi$ varies from $-\pi$ to $\pi$ for $k=\pi$. As a result of this, the factors $(1\pm\cos\Theta(k,\varphi))$  take all values in  $[0, 2]$, which results in the SPSF taking values in $[0, 2\pi]$ at $k=\pi$, for both energy bands in Figs.~\ref{SPSFE}(a,c). On the other hand for a trivial phase in Fig.~\ref{SPSFE}(b), $\Theta(k,\varphi)$ cannot access the whole $[0,\pi]$ interval for any fixed $k$. Specifically, $\Theta(k,\varphi)$ cannot take both $0$ and $\pi$ values at any fixed $k$, as $\varphi$ varies in the BZ. This fact restricts the SPSF in the trivial phase to be either greater than $\pi$ or less than $\pi$ for any fixed $k$, for both upper and lower energy bands, except for two specific $k$ values where it is exactly $\pi$. We find that these features of the SPSF remain same, if we alter the individual values of $g_{\alpha}, g_{\beta}, h$, keeping a constant $g_-/h$ ratio.

\subsection{Qi-Wu-Zhang model}\label{QWZ}
Finally, we consider a 2D topological lattice model. We derive the SPSF for the QWZ model \cite{QWZ2006, Asboth}, a Chern insulator with non-zero Chern numbers. Like the previously discussed 1D topological models, the QWZ model hosts edge states in open boundary conditions in the non-trivial topological phase. The $\textbf{k}$ space Hamiltonian for the QWZ model is,
\begin{align}
H_Q=\sum_{\textbf{k}}\Psi_\textbf{k}^\dagger H_Q(\textbf{k})\Psi_{\textbf{k}},
\end{align}
where, $\Psi_\textbf{k}^\dagger=\begin{pmatrix} \tilde{c}^\dagger_{\textbf{k},\alpha}& \tilde{c}^\dagger_{\textbf{k},\beta}\end{pmatrix}$ and
\begin{align}\label{HQk}
H_{Q}(\textbf{k})&=\begin{pmatrix}({u+\cos{k_x}}+\cos{k_y})&{\sin{k_x}-i\sin{k_y}}\\{\sin{k_x}+i\sin{k_y}}&-({u+\cos{k_x}}+\cos{k_y})\end{pmatrix}\notag\\
&=\vec{d}(\textbf{k})\cdot\vec{\sigma}, 
\end{align}
with the Bloch vector components: $d_x(\textbf{k})=\sin k_x,\quad  d_y(\textbf{k})=\sin k_y$ and $d_z(\textbf{k})=(u+\cos k_x+\cos k_y)$. Here, labels $\alpha$ and $\beta$ indicate two internal states of fermion, e.g., spin, sublattice, etc. The energies of the two bands are $E_{\pm}(\textbf{k})=\pm\sqrt{\sin^2{k_x}+\sin^2{k_y}+({u+\cos{k_x}}+\cos{k_y})^2}=\pm|\vec{d}(\textbf{k})|$. The gap between the bands vanishes when $E_{\pm}(\textbf{k})=0$, which occurs when $|\vec{d}(\textbf{k})|=0$, which is the origin of $d_x$-$d_y$-$d_z$ space. 

\begin{figure*}[t]
  \centering
  {\includegraphics[width=1\textwidth]{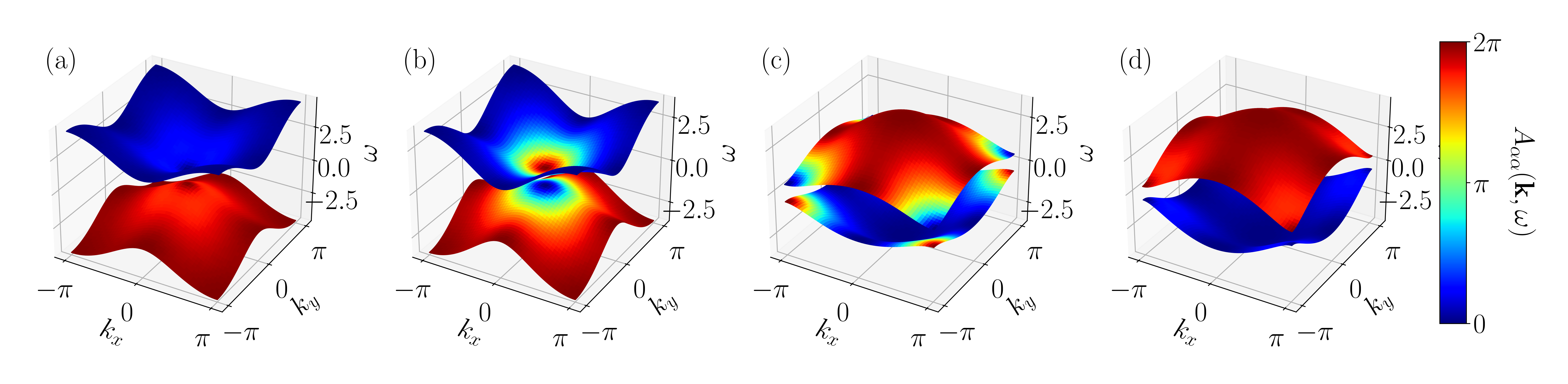}}
  \captionsetup{justification=justified,singlelinecheck=false}
  \caption{Diagonal SPSF $(A_{\alpha\alpha}(\textbf{k},\omega))$ with $\textbf{k},\omega$ for $Q^{-}=0,\pm1$ of the QWZ model. (a) $Q^{-}=0$, $u=-2.5$. The value of SPSF is greater (lesser) than $\pi$ for lower (upper) energy band, in the entire BZ. (b) $Q^{-}=-1$, $u=-1.5$. SPSF has two regions in the BZ, for both the energy bands, with values greater and less than $\pi$, which are separated by a contour of value exactly $\pi$. (c) $Q^{-}=+1$, $u=1.5$. It shows similar features as (b). (d) $Q^{-}=0$, $u=2.5$. The SPSF is smaller (greater) than $\pi$ for lower (upper) energy band, in entire BZ.}
  \label{SPSF_QWZ}
\end{figure*}

Thus, the gap closing can occur at four distinct points of the BZ, which are  (i) $k_x=k_y=0$ if $u=-2$, (ii) $k_x=0,\quad k_y=\pi$ if $u=0$, (iii) $k_x=\pi,\quad k_y=0$ if $u=0$, and (iv) $k_x=k_y=\pi$ if $u=2$. $H_{Q}(\textbf{k})$ provides a map from the 2D BZ, which is a torus, to the space spanned by $d_x, d_y$ and $d_z$. As $\textbf{k}$ sweeps across the 2D BZ, the endpoint of $\vec{d}(\textbf{k})$ traces a deformed torus in the $d_x$-$d_y$-$d_z$ space. The QWZ model has different topological phases with Chern number of the lower band, $Q^{-}=0,\pm 1$. The zero or non-zero value of $Q^{-}$ indicates whether this torus excludes or encloses the origin of $d_x$-$d_y$-$d_z$ space \citep{Asboth}. Again, in terms of the spherical coordinate system $(r, \Theta, \Phi)$ (as in Sec.~\ref{WNCN} and App. \ref{TNT2Dtheta}), the accessible values of $\Theta(\textbf{k})$ reveal the trivial or non-trivial phase, following the arguments similar to those in Sec.~\ref{ExtendedSSH} for the long-range SSH chain.  The phase boundaries separating different Chern numbers are given as \cite{Asboth}:  $Q^{-}=0$ for $2<u$ or $u<-2$, $Q^{-}=-1$ for $-2<u<0$, and $Q^{-}=1$ for $0<u<2$. 

We find that the signature of Chern number in topological phases of the QWZ model can be traced from the diagonal SPSF calculated for the same spin or sublattice (say $\alpha$). So we define a diagonal retarded Green's function in the half-filled many-body ground state $\ket{\mathcal{G}}$ as,
\begin{align}
G^R_{\alpha \alpha }(\textbf{r}_1,t;\textbf{r}_2,0)=-i\theta(t)\bra{\mathcal{G}}\{c_{\textbf{r}_1,\alpha } (t), c^\dagger_{\textbf{r}_2,\alpha }(0)\}\ket{\mathcal{G}}.
\end{align}
We find in App. \ref{AppendixC},
\begin{align}\label{17}
&G^R_{\alpha \alpha }(\textbf{r}_1,t;\textbf{r}_2,0)\notag\\
  &=-i\frac{\theta(t)}{2N_xN_y}\sum_{\textbf{k}}e^{i\textbf{k}\cdot(\textbf{r}_1-\textbf{r}_2)}\Bigl[e^{-iE(\textbf{k})t}\Bigl(1+\cos\Theta(\textbf{k})\Bigr)\notag\\
  &~~~+e^{iE(\textbf{k})t}\Bigl(1-\cos\Theta(\textbf{k})\Bigr)\Bigr],
\end{align}
where, $\cos\Theta(\textbf{k})=d_z(\textbf{k})/|\vec{d}(\textbf{k})|$ and $E(\textbf{k})=|E_{\pm}(\textbf{k})|$. $N_x$ and $N_y$ are the number of unit cells in $x$ and $y$ direction, respectively. 
Since $G_{\alpha \alpha}^R$ in Eq.~\ref{17} is a function of $\textbf{r}_1-\textbf{r}_2$ and $t$ due to space and time translation symmetry, we can take its Fourier transform with respect to $\textbf{r}_1-\textbf{r}_2$ and $t$ to get $G_{\alpha \alpha}^R(\textbf{k},\omega)$ (see App. \ref{AppendixC}). The SPSF from $G_{\alpha \alpha}^R(\textbf{k},\omega)$ is found to be
\begin{align}\label{19}
A_{\alpha \alpha}(\textbf{k},\omega)&=-2\text{Im}G_{\alpha \alpha}^R(\textbf{k},\omega)\notag\\
&=\pi\Bigl(1+\cos\Theta(\textbf{k})\Bigr)\delta(\omega-E(\textbf{k}))\notag\\
  &~~~+\pi\Bigl(1-\cos\Theta(\textbf{k})\Bigr)\delta(\omega+E(\textbf{k})).
\end{align}

This expression is almost identical to that of the long-range SSH chain. We expect the SPSF to show qualitatively different features in different topological phases of the QWZ model due to the presence of $(1\pm\cos\Theta(\textbf{k}))$ factors in Eq.~\ref{19}. In Fig.~\ref{SPSF_QWZ}, we show the color plots of SPSF with $\textbf{k},\omega$ for $Q^{-}=0,\pm1$. We observe a clear qualitative difference in the behavior of $A_{\alpha \alpha}(\textbf{k},\omega)$ for zero and non-zero Chern numbers. Figs.~\ref{SPSF_QWZ}(a,d) show that $A_{\alpha \alpha}(\textbf{k},\omega)$ is either greater (less) than $\pi$ for $u<-2$ or less (greater) than $\pi$ for $u>2$, for lower (upper) energy band in the entire BZ, when the model is in a trivial phase ($Q^{-}=0$). As $k_x$ varies from $-\pi$ to $\pi$ for any specific $k_y$, $\Theta(\textbf{k})$ is always either between $0$ and $\pi/2$ (when $u>2$) or between $\pi/2$ and $\pi$ (when $u<-2$).  Thus, the value of $A_{\alpha \alpha}(\textbf{k},\omega)$ remains either lesser or greater than $\pi$, depending on whether we are looking at upper or lower energy band, and whether $u>2$ or $u<-2$. On the contrary, when the system is in a non-trivial phase for $Q^{-}=\pm 1$, there are two regions of the SPSF, for both energy bands in the BZ in Figs.~\ref{SPSF_QWZ}(b,c), where $A_{\alpha \alpha}(\textbf{k},\omega)$ is greater than $\pi$ and less than $\pi$, and these regions are separated by a contour where it is exactly $\pi$. Therefore, the value of SPSF becomes $\pi$ twice as $k_x~({\rm or}~ k_y)$ varies from $-\pi$ to $\pi$ for $k_y=0~({\rm or}~ k_x=0)$. It is again, purely the consequence of topology for reasons similar to those given in the previous sub-section. The above features of the SPSF remain robust when we change the parameters within each topological phase of the QWZ model. Thus, the SPSF can be applied to decisively distinguish the trivial and non-trivial phases of the QWZ model.

Interestingly, when we change a Bloch vector component, e.g., $d_y(\textbf{k})=\sin k_y$ to $d_y(\textbf{k})=\sin^2 k_y$ in Eq. \ref{HQk}, the model is no longer topological, yet the $\Theta(\textbf{k})$ behavior remains unchanged. So, the SPSF may give false positive results for such topologically trivial models since it depends only on $\Theta(\textbf{k})$. This is a limitation of our method, which has to be complemented with another measurement method to confirm the non-trivial topological nature of the matter under investigation. The non-trivial topology, i.e., a non-zero Chern number, can be detected by probing the edge states of topological materials (satisfying bulk-boundary correspondence)  using STS or ARPES, which can also probe local SPSF in real space. We now show a method using first-order coherence, which can be used in complement with SPSF. The first-order coherence is related to the lesser Green's function, which is defined as
\begin{align}\label{FOC}
C_{\alpha \beta}(\textbf{r}_1-\textbf{r}_2,t)=i\bra{\mathcal{G}}c_{\textbf{r}_1, \alpha}^\dagger (t)c_{\textbf{r}_2, \beta}(0)\ket{\mathcal{G}}.
\end{align}
Its Fourier transform in $\textbf{r}_1-\textbf{r}_2$ and $t$ gives us first-order coherence in $\textbf{k}$ and $\omega$ space as
\begin{align}
C_{\alpha \beta}(\textbf{k},\omega)=-i\pi\sin\Theta(\textbf{k})e^{-i\Phi(\textbf{k})}\delta(\omega-E(\textbf{k})).
\end{align}
Since it contains $\Theta(\textbf{k})$ and $\Phi(\textbf{k})$, it can be used along with the SPSF to measure the bulk  Chern number of 2D topological matter. For the model with $d_y(\textbf{k})=\sin^2k_y$, the real part of $C_{\alpha \beta}(\textbf{k},\omega)$ can be applied to probe it as a trivial insulator. First-order coherence can be probed using interferometer set-ups \cite{Damm2017}. 

\section{Summary and outlook}\label{sum}
In conclusion, we propose a method to directly probe winding and Chern numbers using the momentum-space spectral function in 1D and 2D topological materials. Our proposal is attainable within the current scope of various ARPES and STS measurements. Here, we have derived the momentum-space SPSF for clean systems and related it to the winding or Chern numbers of the systems. The location of Wannier centers in artificial 1D lattices made of Cs atoms on InAs is experimentally found by integrating the spatially resolved local density of states in a recent study with a scanning tunneling microscope \cite{Ligthart2024}. The position of the Wannier centers can give the bulk topological properties for various 1D chains in trivial and non-trivial phases when topological invariants, such as the winding number, are not directly experimentally accessible. One immediate obstacle to our proposal can be imperfection or disorder in the parameters of the natural materials. We tested the disorder in various parameters, e.g., hopping amplitudes or onsite energy of the model. Our main results of acquiring knowledge of bulk topological invariants from the SPSF remain valid even for a substantial amount of randomness. We have demonstrated it for the Kitaev model in App.~\ref{Disorder}.

For the long-range SSH chain without chiral symmetry and the QWZ model, we have shown that the diagonal SPSF in sublattice basis or internal states can identify the bulk topology of the model. This requires a high-resolution spin-resolved or sublattice-resolved ARPES or STS setup. A spin-resolved ARPES setup has been developed using the time of flight technique to visualize spin textures in electronic band structures \cite{Hsieh2009,Jozwiak2011,Zhu2013,Jozwiak2016}. These experiments were carried out using high-efficiency and high-resolution spin-resolved photo-electron spectrometers. Similar high-efficiency and high-resolution ARPES measurements by selectively probing one particular sublattice or spin state can detect our proposed diagonal SPSF. 

Here, we consider that the ARPES essentially measures the SPSF of the initial states of the unperturbed system, which is true in the first approximation and within the so-called three-step model \cite{Bansil1999}. Nevertheless, a more accurate description of the measured spectra from ARPES requires better modeling of the photoexcitation process by adequately taking into account the matrix element of the light-matter interaction Hamiltonian between initial and final states of the matter \cite{Bansil1999}. While the SPSF determines the existence of a peak, its intrinsic intensity, and its width in the measured ARPES spectra, the matrix element modulates intrinsic intensities according to the geometric experimental constraints. Overall, the ARPES data depend on the experimental details. Thus, further work within the so-called one-step model \cite{Bansil1999} is required to refine the connection between the measured photocurrent from ARPES and the winding and Chern numbers of topological matter. Our work is essential to this goal by showing a direct connection between the SPSF and winding and Chern numbers of some topological matters. 


The SPSF works well for detecting winding numbers of 1D topological matter. For 2D topological matters, the Chern number depends on both $\Theta(\textbf{k})$ and $\Phi(\textbf{k})$. Our calculated SPSF for different topological matters contains only $\Theta(\textbf{k})$. Nevertheless, $\Theta(\textbf{k})$ has different features in the trivial and non-trivial phases of various well-known 2D topological models, which we explored in this paper. In App.~\ref{Haldane}, we investigate another 2D topological insulator, the Haldane model, on the honeycomb lattice. The SPSF behaves similarly for the Haldane model as for the other studied topological matters showing signatures of topology. However, because of only $\Theta(\textbf{k})$ dependence, the SPSF may sometimes give false positive or ambiguous results for models like the one discussed in the paragraph before Eq. \ref{FOC} and has to be used in complement with other measurements. Hence, the SPSF-based detection must be complemented by another measurement to ensure the matter under investigation is topological, i.e., having a non-zero Chern number. We also propose another method with the first-order coherence, which contains $\Theta(\textbf{k})$ and $\Phi(\textbf{k})$ information and can be used along with the SPSF to determine the Chern number.  

Extending the discussion from 1D and 2D two-band models to multi-band models and three-dimensional topological materials would be exciting. In App.~\ref{Fourband}, we study a four-band model, which was proposed as an experimental realization of the Kitaev chain. The SPSF in this four-band model, too, is shown to have signatures of trivial and non-trivial phases. In recent years, massive growth has been witnessed in the study of topology in dissipative systems with non-Hermitian Hamiltonians. We tried to derive SPSF in non-Hermitian systems, such as non-Hermitian SSH models with non-reciprocal hopping \cite{Vyas21, ritu}. The SPSF is complex-valued for this case. It seems that both the real and imaginary parts of the SPSF are relevant in detecting the bulk topology of such dissipative systems. We plan to pursue a systematic study in this direction shortly.

\section{Acknowledgements}
We thank Arun Paramekanti for a helpful discussion. KBE thanks Kshitij Sharma for useful discussions.

\appendix
\section{Trivial vs. non-trivial topology of 2D topological insulators}\label{TNT2Dtheta}
\begin{figure}[h]
  \centering
  {\includegraphics[width=0.4\textwidth]{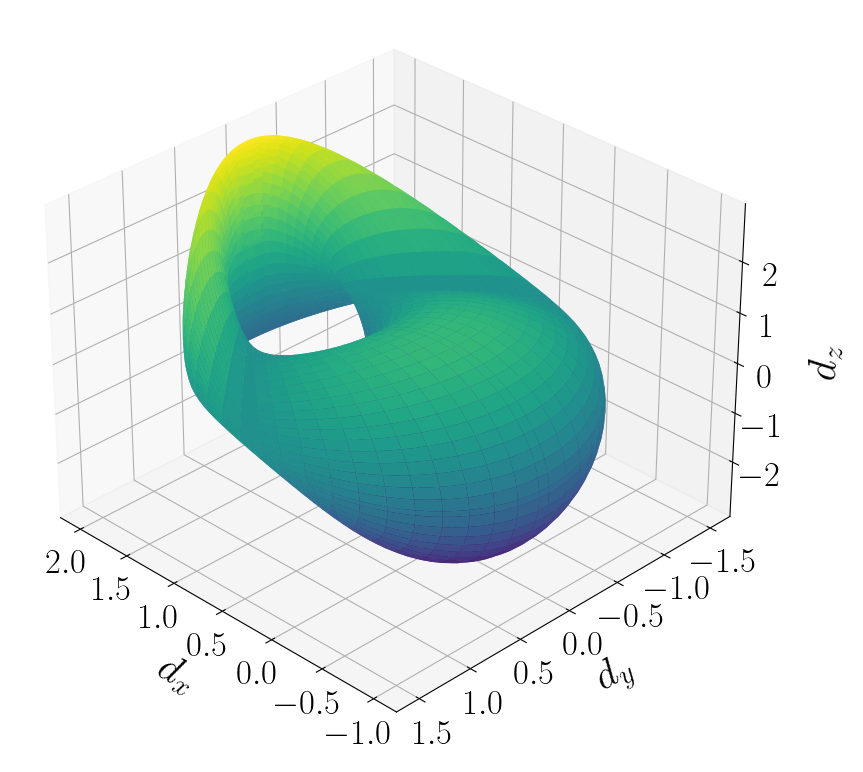}}
  \caption{Deformed torus traced by the tip of $\vec{d}(k,\varphi)$ of the long-range SSH chain in $d_x(\textbf{k})$-$d_y(\textbf{k})$-$d_z(\textbf{k})$ space as $k$ and $\varphi$ traverse the effective 2D BZ.}
  \label{defTLRSSH}
\end{figure}

In this appendix, we discuss the trivial vs non-trivial topology of a generic 2D topological insulator described by a Bloch vector $\vec{d}(\textbf{k})$. 
The Hamiltonian $H(\textbf{k})$ gives a map from the 2D BZ, which is a torus, to the $d_x(\textbf{k})$-$d_y(\textbf{k})$-$d_z(\textbf{k})$ space. As the momenta $k_x$ and $k_y$ traverse the 2D BZ, the tip of the Bloch vector $\vec{d}(\textbf{k})$ traces a deformed torus in the $d_x(\textbf{k})$-$d_y(\textbf{k})$-$d_z(\textbf{k})$ space. Fig. \ref{defTLRSSH} shows such a deformed torus for the long-range SSH chain. The origin of this space corresponds to the band gap closing. If the origin lies outside (inside) the torus, the insulator is said to be in a trivial (non-trivial) phase.

\begin{figure}[h]
  \centering
  \subfloat[]{\includegraphics[width=0.23\textwidth]{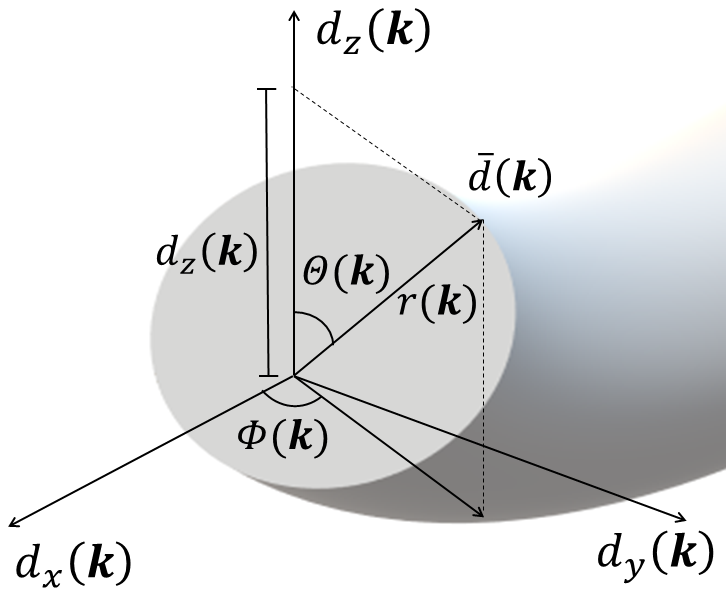}}
  \hfill
  \subfloat[]{\includegraphics[width=0.23\textwidth]{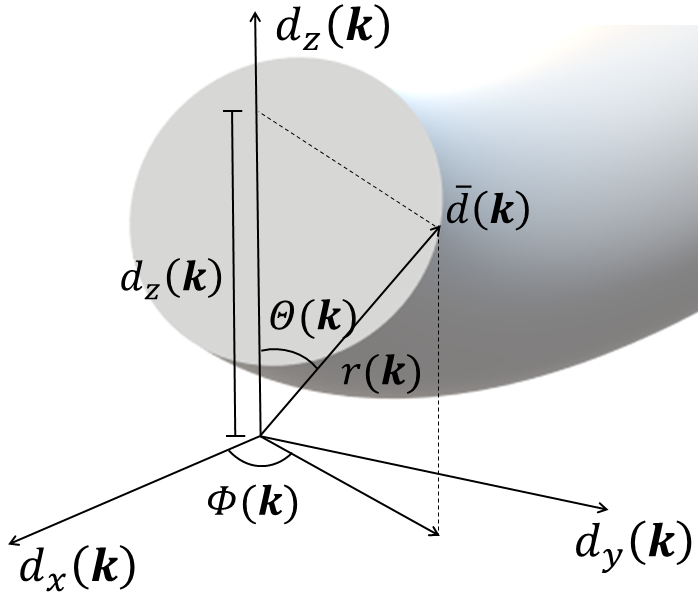}}
  \hfill
  \subfloat[]{\includegraphics[width=0.23\textwidth]{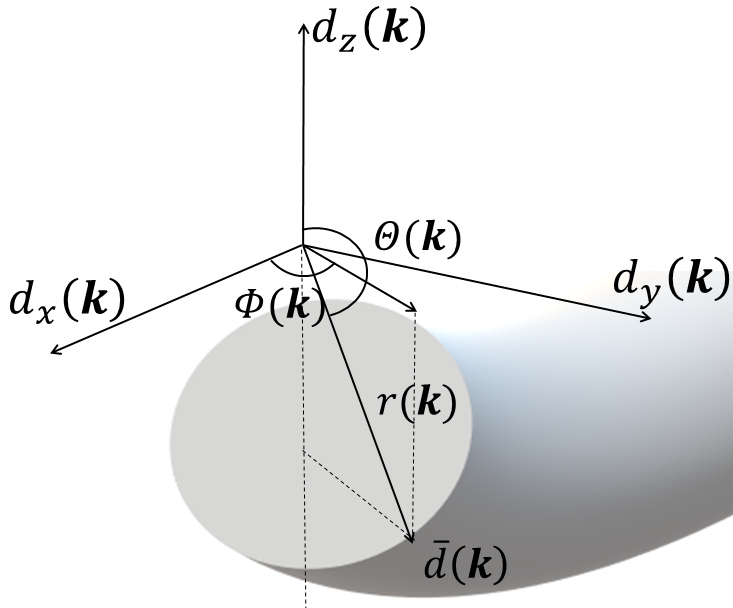}}
  \hfill
  \subfloat[]{\includegraphics[width=0.23\textwidth]{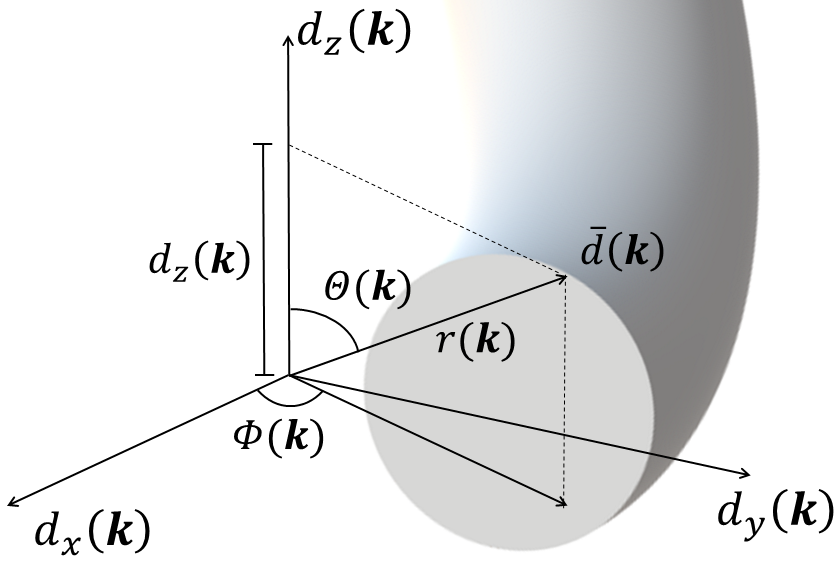}}
  \caption{Cross-sections of the tori for various scenarios, describing non-trivial phase in (a) and trivial phase in (b), (c) and (d). It can be understood that in plot (a), $\Theta(\textbf{k})$ varies from $0$ to $\pi$. In plots (b) and (c), $\Theta(\textbf{k})$ varies in the range $[0,\tilde{\Theta}_0]$ and $[\pi-\tilde{\Theta '}_0,\pi]$, respectively, where $\tilde{\Theta}_0, \tilde{\Theta '}_0 <\pi/2$. In plot (d), $\Theta(\textbf{k})$ varies between $\Theta_0$ and $\pi-\Theta_0$, where $\Theta_0<\pi/2$. }
  \label{2DTNT}
\end{figure}

Fig. \ref{2DTNT} displays different cross-sections of the tori at a fixed $k_x$ or $k_y$, say $k_x=k_0$ for different cases, e.g., trivial or non-trivial topology. Each point on the torus can be described by the spherical polar coordinates $(r(\textbf{k}), \Theta(\textbf{k}),\Phi(\textbf{k}))$, where $r(\textbf{k})=|\vec{d}(\textbf{k})|$, $\Theta(\textbf{k})$ is the angle between $\vec{d}(\textbf{k})$ and the $d_z(\textbf{k})$ axis, and $\Phi(\textbf{k})$ is the angle between the projection of $\vec{d}(\textbf{k})$ on the $d_x(\textbf{k})$-$d_y(\textbf{k})$ plane and the $d_x(\textbf{k})$ axis. Fig. \ref{2DTNT}(a) shows the situation for a non-trivial phase. The origin is enclosed by the torus, and it can be understood from the figure that the angle $\Theta(\textbf{k})$ takes all values in $[0,\pi]$ as $k_x$ is fixed at $k_0$, and $k_y$ varies in the BZ. Nevertheless, the angle $\Theta(\textbf{k})$ cannot access the full $[0,\pi]$ range in the trivial phase. Instead, $\Theta(\textbf{k})$ can either be in $[0,\pi/2)$, $(\pi/2,\pi]$ or $[\Theta_0,\pi-\Theta_0]$, where $\Theta_0<\pi/2$. It can take values in any other proper subinterval of $[0,\pi]$, but never the full $[0,\pi]$ interval, for any fixed $k_x$. For illustration, we take three scenarios of the trivial phase in Figs.~\ref{2DTNT}(b), (c) and (d). Figs.~\ref{2DTNT}(b) and (c) display the case when $\Theta(\textbf{k})$ lies in the range $[0,\pi/2)$ and $(\pi/2,\pi]$ respectively. Fig.~\ref{2DTNT}(d) illustrates the case when $\Theta(\textbf{k})$ lies in the range $[\Theta_0,\pi-\Theta_0]$, where $\Theta_0<\pi/2$. It is clear that the torus in Fig.~\ref{2DTNT}(a) cannot be adiabatically (continuously) deformed into any of the tori in Figs.~\ref{2DTNT}(b,c,d), without passing through the origin, hence closing the band gap. Thus, these represent topologically distinct phases.

\section{Derivation of SPSF for Kitaev chain} \label{AppendixB}
To derive the retarded Green's function and the SPSF for the Kitaev chain, we start with the diagonalized Hamiltonian in Eq. \ref{dHK}.
\begin{align}\label{B1}
	  H_{K}=\sum_{k}\Bigl(E(k)\tilde{\eta}_{k}^\dagger \tilde{\eta}_{k} -E(-k)\tilde{\eta}_{-k}\tilde{\eta}_{-k}^\dagger \Bigr),
\end{align}
where, $E(k)=\sqrt{\Delta^2 \sin^2k+(\mu/2+J\cos{k})^2}$ and $\tilde{\eta}_{k}$ are the Bogoliubov excitations given by,
\begin{align}\label{B2}
	\tilde{\eta}_{k}=X(k)\tilde{c}_k+Y^*(k)\tilde{c}_{-k}^\dagger,
\end{align}
where, $X(k)=\cos\tilde{\theta}(k)/\sqrt{2(1-\sin\tilde{\theta}(k))}$ and $Y(k)=i\sqrt{(1-\sin\tilde{\theta}(k))/2}$.
The angle $\tilde{\theta}(k)$ is defined in the paragraph before Eq. \ref{dHK}.\\

The ground state of the Hamiltonian, $\ket{\mathcal{G}}$ is the vacuum of the Bogoliubov quasiparticles, which is defined as,
\begin{align}\label{B4}
	\tilde{\eta}_k\ket{\mathcal{G}}=0, \quad \forall k \in BZ .
\end{align}
Now, we derive the retarded Green's function and the SPSF for the Bogoliubov vacuum. 
The retarded Green's function at zero temperature is defined as,
\begin{align}\label{B5}
	G^R(x_1,t;x_2,0)&=-i\theta(t)\bra{\mathcal{G}}\{c_{x_1}(t), c_{x_2}^\dagger(0)\}\ket{\mathcal{G}}\notag\\
	&=-i\theta(t)\Bigl(\bra{\mathcal{G}}c_{x_1}(t) c_{x_2}^\dagger(0)\ket{\mathcal{G}}\notag\\
	&+\bra{\mathcal{G}} c_{x_2}^\dagger(0)c_{x_1}(t)\ket{\mathcal{G}}\Bigr).
\end{align}
Now, we separately calculate $\bra{\mathcal{G}}c_{x_1}(t) c_{x_2}^\dagger(0)\ket{\mathcal{G}}$ and $\bra{\mathcal{G}} c_{x_2}^\dagger(0)c_{x_1}(t)\ket{\mathcal{G}}$. Lets do $\bra{\mathcal{G}}c_{x_1}(t) c_{x_2}^\dagger(0)\ket{\mathcal{G}}$ first. 
Applying PBC and taking Fourier transform of the creation and annihilation operators, e.g.,
\begin{align}
	c_x=\frac{1}{\sqrt{N}}\sum_ke^{ikx}\tilde{c}_k,
\end{align}
where, $N$ is the system size, we get,
\begin{align}\label{B6}
	\bra{\mathcal{G}}c_{x_1}(t) c_{x_2}^\dagger(0)\ket{\mathcal{G}}=\frac{1}{N}\sum_{k,l}e^{i(x_1k-x_2l)}\bra{\mathcal{G}}\tilde{c}_k(t)\tilde{c}_l^\dagger(0)\ket{\mathcal{G}}.
\end{align}
To find $\tilde{c}_k(t)$, we obtain its differential equation using Heisenberg equation,  
\begin{align}\label{B7}
	\dot{\tilde{c}}_k(t)=i(\mu/2+J\cos k)\tilde{c}_k(t)-(\Delta \sin k) \tilde{c}_{-k}^\dagger(t).
\end{align}
Similarly, we obtain the equation for $\tilde{c}_{-k}^\dagger(t)$.
\begin{align}\label{B8}
	\dot{\tilde{c}}_{-k}^\dagger(t)=-i(\mu/2+J\cos k)\tilde{c}_{-k}^\dagger(t)+(\Delta \sin k) \tilde{c}_k(t).
\end{align}
Eqs. \ref{B7} and \ref{B8} are coupled differential equations for $\tilde{c}_k(t)$ and $\tilde{c}_{-k}^\dagger(t)$. Solving for $\tilde{c}_k(t)$, and by taking initial conditions $\tilde{c}_k(t=0)=\tilde{c}_k(0)$ and $\tilde{c}_{-k}^\dagger(t=0)=\tilde{c}_{-k}^\dagger(0)$, we get,
\begin{align}\label{B9}
	\tilde{c}_k(t)=\tilde{P}^*(k,t)\tilde{c}_k(0)+\tilde{Q}^*(k,t)\tilde{c}_{-k}^\dagger(0),
\end{align}
where, $\tilde{P}(k,t)=i\sin\tilde{\theta}(k)\sin E(k)t+\cos E(k)t$ and $\tilde{Q}(k,t)=-\cos\tilde{\theta}(k)\sin E(k)t$.

Substituting this in $\bra{\mathcal{G}}c_{k}(t) c_{l}^\dagger(0)\ket{\mathcal{G}}$, we get,
\begin{align}\label{B11}
	\bra{\mathcal{G}}c_{k}(t) c_{l}^\dagger(0)\ket{\mathcal{G}}=\tilde{P}^*(k,t)\bra{\mathcal{G}}\tilde{c}_k(0)\tilde{c}_l^\dagger(0)\ket{\mathcal{G}}\notag\\
	+\tilde{Q}^*(k,t)\bra{\mathcal{G}}\tilde{c}_{-k}^\dagger(0)\tilde{c}_l^\dagger(0)\ket{\mathcal{G}}.
\end{align}
Further, $\bra{\mathcal{G}}\tilde{c}_k(0)\tilde{c}_l^\dagger(0)\ket{\mathcal{G}}$ and $\bra{\mathcal{G}}\tilde{c}_{-k}^\dagger(0)\tilde{c}_l^\dagger(0)\ket{\mathcal{G}}$ can be calculated using Eqs. \ref{B2} and \ref{B4}. They come out to be,
\begin{align}\label{B12}
	\bra{\mathcal{G}}\tilde{c}_k(0)\tilde{c}_l^\dagger(0)\ket{\mathcal{G}}&=\frac{\cos^2\tilde{\theta}(k)}{2(1-\sin\tilde{\theta}(k))}\delta_{k,l},\notag\\
	\bra{\mathcal{G}}\tilde{c}_{-k}^\dagger(0)\tilde{c}_l^\dagger(0)\ket{\mathcal{G}}&=\frac{i\cos\tilde{\theta}(k)}{2}\delta_{k,l}.
\end{align}
Putting these back in $\bra{\mathcal{G}}c_{k}(t) c_{l}^\dagger(0)\ket{\mathcal{G}}$, and substituting the expressions for $\tilde{P}^*(k,t)$ and $\tilde{Q}^*(k,t)$, we get

\begin{align}
	\bra{\mathcal{G}}c_{k}(t) c_{l}^\dagger(0)\ket{\mathcal{G}}=\frac{1+\sin\tilde{\theta}(k)}{2}e^{-iE(k)t}\delta_{k,l}.
\end{align}

Using this in \ref{B6}, and performing the summation over $l$, we get

\begin{align}
	\bra{\mathcal{G}}c_{x_1}(t) c_{x_2}^\dagger(0)\ket{\mathcal{G}}=\frac{1}{2N}\sum_{k}e^{ik(x_1-x_2)}e^{-iE(k)t}\Bigl(1+\sin\tilde{\theta}(k)\Bigr).
\end{align}
Similarly, we calculate $\bra{\mathcal{G}}c_{x_2}^\dagger(0)c_{x_1}(t) \ket{\mathcal{G}}$. It comes out to be

\begin{align}
	\bra{\mathcal{G}}c_{x_2}^\dagger(0)c_{x_1}(t) \ket{\mathcal{G}}=\frac{1}{2N}\sum_{k}e^{ik(x_1-x_2)}e^{iE(k)t}\Bigl(1-\sin\tilde{\theta}(k)\Bigr).
\end{align}

Putting everything together, we finally get the retarded Green's function,
\begin{align}\label{B13}
	G^R(x_1,t;x_2,0)&=-i\frac{\theta(t)}{2N}\sum_k e^{ik(x_1-x_2)}\Bigl[e^{-iE(k)t}\Bigl(1+\sin\tilde{\theta}(k)\Bigr)\notag\\
	&+e^{iE(k)t}\Bigl(1-\sin\tilde{\theta}(k)\Bigr)\Bigr].
\end{align}
Its Fourier transform in $(x_1-x_2)\equiv r$ and $t$ is given by,
\begin{align}\label{B14}
G^R(k,\omega)=\int _{-\infty}^{\infty} dt e^{i\omega t}\sum_{r=0}^N  e^{ikr}G^R(x_1,t;x_2,0).
\end{align}
The Fourier transform in $r$ gives us 
\begin{align}
\sum_{r=0}^N  e^{ikr}G^R(x_1,t;x_2,0)&=-i\frac{\theta(t)}{2}\Bigl[e^{-iE(k)t}\Bigl(1+\sin\tilde{\theta}(k)\Bigr)\notag\\
	&+e^{iE(k)t}\Bigl(1-\sin\tilde{\theta}(k)\Bigr)\Bigr].
\end{align}
We notice that the retarded Green's function is not decaying at long time. Therefore, for the time Fourier transform to converge, we need to replace the real frequency $\omega$ by $\omega+i\epsilon$, and take the limit $\epsilon \rightarrow 0^+$ in the end, where $\epsilon$ is a positive real number. This gives us
\begin{align}
G^R(k,\omega)&=\frac{-i}{2}\int _{-\infty}^{\infty} dt e^{i(\omega+i\epsilon) t}\theta(t)\Bigl[e^{-iE(k)t}\Bigl(1+\sin\tilde{\theta}(k)\Bigr)\notag\\
	&+e^{iE(k)t}\Bigl(1-\sin\tilde{\theta}(k)\Bigr)\Bigr]\notag\\
	&=\frac{-i}{2}\Bigl[\Bigl(1+\sin\tilde{\theta}(k)\Bigr)\int _{0}^{\infty} dt e^{i(\omega -E(k)+i\epsilon)t}\notag\\
	&+\Bigl(1-\sin\tilde{\theta}(k)\Bigr)\int _{0}^{\infty} dt e^{i(\omega +E(k)+i\epsilon)t}\Bigr]\notag\\
	&=\frac{1}{2}\Bigl[\frac{1+\sin\tilde{\theta}(k)}{\omega -E(k)+i\epsilon}+\frac{1-\sin\tilde{\theta}(k)}{\omega +E(k)+i\epsilon}\Bigr].
\end{align}
Now, we take the limit $\epsilon \rightarrow 0^+$ and use the result 
\begin{align}\label{CPV}
\lim_{\epsilon \rightarrow 0^+}\frac{1}{x\pm i\epsilon}=\mathcal{P}\Big(\frac{1}{x}\Bigr)\mp i\pi\delta(x),
\end{align}
where, $\mathcal{P}$ denotes the Cauchy principal value.
\begin{align}
	&G^R(k,\omega)\notag\\
	&=\frac{1}{2}\Bigl[\bigl(1+\sin\tilde{\theta}(k)\bigr)\Bigl(\mathcal{P}\Big(\frac{1}{\omega -E(k)}\Bigr)- i\pi\delta(\omega -E(k))\Bigr)\notag\\
	&+\bigl(1-\sin\tilde{\theta}(k)\bigr)\Bigl(\mathcal{P}\Big(\frac{1}{\omega +E(k)}\Bigr)- i\pi\delta(\omega +E(k))\Bigr)\Bigr].
\end{align}
This gives us the SPSF as 
\begin{align}
	A(k,\omega)&=-2\text{Im}G^R(k,\omega)\notag\\
	&=\pi(1+\sin\tilde{\theta}(k))\delta(\omega-E(k))\notag\\
	&+\pi(1-\sin\tilde{\theta}(k))\delta(\omega+E(k)).\label{sumR}
\end{align}

The spectral function has to satisfy a summation rule \cite{Bruus2004} given by
\begin{align}\label{sumrule}
\frac{1}{2\pi}\int _{-\infty}^{\infty}  A(\textbf{k},\omega)d\omega=1.
\end{align}

$A(k,\omega)$ in Eq.~\ref{sumR} for the Kitaev chain satisfies the above sum rule.
\begin{align}
\frac{1}{2\pi}\int _{-\infty}^{\infty} A(k,\omega)d\omega&=\frac{1}{2}(1+\sin\tilde{\theta}(k))\int _{-\infty}^{\infty} \delta(\omega-E(k))d\omega\notag\\
&+\frac{1}{2}(1-\sin\tilde{\theta}(k))\int _{-\infty}^{\infty} \delta(\omega+E(k))d\omega\notag\\
&=\frac{1}{2}(1+\sin\tilde{\theta}(k))+\frac{1}{2}(1-\sin\tilde{\theta}(k))\notag\\
&=1.
\end{align}

\section{Derivation of SPSF for QWZ model}\label{AppendixC}

In this appendix, we derive the retarded Green's function and the SPSF for the QWZ model at zero temperature. Lets start with the $k$ space Hamiltonian in Eq. \ref{HQk},
\begin{align}
H_{Q}(\textbf{k})&=\begin{pmatrix}({u+\cos{k_x}}+\cos{k_y})&{\sin{k_x}-i\sin{k_y}}\\{\sin{k_x}+i\sin{k_y}}&-({u+\cos{k_x}}+\cos{k_y})\end{pmatrix}\notag\\
&=\vec{d}(\textbf{k})\cdot\vec{\sigma}, 
\end{align}
where, the Bloch vector has components: $d_x(\textbf{k})=\sin k_x,\quad  d_y(\textbf{k})=\sin k_y$ and $d_z(\textbf{k})=(u+\cos k_x+\cos k_y)$. The energies of the two bands are $E_{\pm}(\textbf{k})=\pm\sqrt{\sin^2{k_x}+\sin^2{k_y}+({u+\cos{k_x}}+\cos{k_y})^2}=\pm|\vec{d}(\textbf{k})|$. The diagonalized Hamiltonian in terms of the normal mode creation-annihilation operators is given by
\begin{align}
     H_{Q}=\sum_\textbf{k}\Bigl({E(\textbf{k})} \tilde{\zeta}_{\textbf{k},+}^\dagger \tilde{\zeta}_{\textbf{k},+}-{E(\textbf{k})} \tilde{\zeta}_{\textbf{k},-}^\dagger \tilde{\zeta}_{\textbf{k},-}\Bigr),
\end{align}
where, $E(\textbf{k})=|E_{\pm}(\textbf{k})|$ and the relations between $\tilde{c}_{\textbf{k},\alpha/\beta}$ and the normal mode operators $\tilde{\zeta}_{\textbf{k},\pm}$ are given by

\begin{align}\label{czrelation}
     \tilde{c}_{\textbf{k},\alpha}&= \sqrt{\frac{1+\cos\Theta(\textbf{k})}{2}}\tilde{\zeta}_{\textbf{k},+} +\sqrt{\frac{1-\cos\Theta(\textbf{k})}{2}}\tilde{\zeta}_{\textbf{k},-},\notag\\
     \tilde{c}_{\textbf{k},\beta}&= e^{i\Phi(\textbf{k})}\Bigl(\sqrt{\frac{1-\cos\Theta(\textbf{k})}{2}}\tilde{\zeta}_{\textbf{k},+} -\sqrt{\frac{1+\cos\Theta(\textbf{k})}{2}}\tilde{\zeta}_{\textbf{k},-}\Bigr),
\end{align} 
where, $\Theta(\textbf{k})$ and $\Phi(\textbf{k})$ are, respectively, the polar and the azimuthal angles of the vector $\vec{d}(\textbf{k})$ in the spherical polar coordinate system. Even though the relations in Eq.~\ref{czrelation} are gauge-dependent, it is easy to check that the retarded Green's function and the SPSF are independent of the gauge choice. The normal mode operators are fermionic, and satisfy fermionic anti-commutation relations:
\begin{align}
    \{ \tilde{\zeta}_{\textbf{k},\alpha}, \tilde{\zeta}_{\textbf{k}',\beta}^\dagger\}=\delta_{\textbf{k},\textbf{k}'}\delta_{\alpha,\beta}, \qquad \{ \tilde{\zeta}_{\textbf{k},\alpha}, \tilde{\zeta}_{\textbf{k}',\beta}\}=0.
\end{align}

The half filled ground state of this Hamiltonian is given by 
\begin{align}
\ket{\mathcal{G}}=\prod_{\textbf{k}\in BZ}\tilde{\zeta}_{\textbf{k},-}^\dagger\ket{0},
\end{align}
where, $\ket{0}$ is the vacuum state with no particles. The zero temperature diagonal retarded Green's function, for same internal state of spin or sublattice, say $\alpha$, is defined as
\begin{align}
	G_{\alpha \alpha}^R(\textbf{r}_1,t;\textbf{r}_2,0)&=-i\theta(t)\bra{\mathcal{G}}\{c_{\textbf{r}_1,\alpha}(t), c_{\textbf{r}_2,\alpha}^\dagger(0)\}\ket{\mathcal{G}}\notag\\
	&=-i\theta(t)\Bigl(\bra{\mathcal{G}}c_{\textbf{r}_1,\alpha}(t) c_{\textbf{r}_2,\alpha}^\dagger(0)\ket{\mathcal{G}}\notag\\
	&~~~+\bra{\mathcal{G}}c_{\textbf{r}_2,\alpha}^\dagger(0)c_{\textbf{r}_1,\alpha}(t)\ket{\mathcal{G}}\Bigr).
\end{align}
We separately calculate $\bra{\mathcal{G}}c_{\textbf{r}_1,\alpha}(t) c_{\textbf{r}_2,\alpha}^\dagger(0)\ket{\mathcal{G}}$ and $\bra{\mathcal{G}}c_{\textbf{r}_2,\alpha}^\dagger(0)c_{\textbf{r}_1,\alpha}(t)\ket{\mathcal{G}}$. Lets do $\bra{\mathcal{G}}c_{\textbf{r}_1,\alpha}(t) c_{\textbf{r}_2,\alpha}^\dagger(0)\ket{\mathcal{G}}$ first. 
Applying PBC and taking Fourier transform of the creation and annihilation operators,
\begin{align}
	c_{\textbf{r},\alpha}=\frac{1}{\sqrt{N_x}\sqrt{N_y}}\sum_\textbf{k} e^{i\textbf{k}\cdot\textbf{r}}\tilde{c}_{\textbf{k},\alpha},
\end{align}
where, $N_x$ and $N_y$ are the number of unit cells in $x$ and $y$ direction respectively, we get,
\begin{align}\label{D8}
	\bra{\mathcal{G}}c_{\textbf{r}_1,\alpha}(t) c_{\textbf{r}_2,\alpha}^\dagger(0)\ket{\mathcal{G}}=&\frac{1}{N_xN_y}\sum_{\textbf{k},\textbf{l}}e^{i(\textbf{k}\cdot\textbf{r}_1-\textbf{l}\cdot\textbf{r}_2)}\notag\\
	&\cross\bra{\mathcal{G}}c_{\textbf{k},\alpha}(t) c_{\textbf{l},\alpha}^\dagger(0)\ket{\mathcal{G}}.
\end{align}
We can find $c_{\textbf{k},\alpha}(t)$ in a manner similar to that used in the previous appendix, by employing the Heisenberg equation:
\begin{align}
c_{\textbf{k},\alpha}(t)&=P^*(\textbf{k},t)c_{\textbf{k},\alpha}(0)
+Q^*(\textbf{k},t)c_{\textbf{k},\beta}(0),
\end{align}
where, $P(\textbf{k},t)=\cos E(\textbf{k})t+i\cos\Theta(\textbf{k})\sin E(\textbf{k})t$ and $Q(\textbf{k},t)=ie^{i\Phi(\textbf{k})}\sin\Theta(\textbf{k})\sin E(\textbf{k})t$.
Using this, $\bra{\mathcal{G}}c_{\textbf{k},\alpha}(t) c_{\textbf{l},\alpha}^\dagger(0)\ket{\mathcal{G}}$ becomes
\begin{align}\label{D10}
\bra{\mathcal{G}}c_{\textbf{k},\alpha}(t) c_{\textbf{l},\alpha}^\dagger(0)\ket{\mathcal{G}}=P^*(\textbf{k},t)\bra{\mathcal{G}}c_{\textbf{k},\alpha}(0)c_{\textbf{l},\alpha}^\dagger(0)\ket{\mathcal{G}}\notag\\
+Q^*(\textbf{k},t)\bra{\mathcal{G}}c_{\textbf{k},\beta}(0)c_{\textbf{l},\alpha}^\dagger(0)\ket{\mathcal{G}}.
\end{align}
The terms $\bra{\mathcal{G}}c_{\textbf{k},\alpha}(0)c_{\textbf{l},\alpha}^\dagger(0)\ket{\mathcal{G}}$ and $\bra{\mathcal{G}}c_{\textbf{k},\beta}(0)c_{\textbf{l},\alpha}^\dagger(0)\ket{\mathcal{G}}$ can be worked out using \ref{czrelation}.
\begin{align}
\bra{\mathcal{G}}c_{\textbf{k},\alpha}(0)c_{\textbf{l},\alpha}^\dagger(0)\ket{\mathcal{G}}&=\frac{1+\cos\Theta(\textbf{k})}{2}\delta_{\textbf{k},\textbf{l}},\notag\\
\bra{\mathcal{G}}c_{\textbf{k},\beta}(0)c_{\textbf{l},\alpha}^\dagger(0)\ket{\mathcal{G}}&=\frac{e^{i\Phi(\textbf{k})}}{2}\sin\Theta(\textbf{k})\delta_{\textbf{k},\textbf{l}}.
\end{align}
Putting these back in \ref{D10}, and using the expressions for $P^*(\textbf{k},t)$ and $Q^*(\textbf{k},t)$, we get
\begin{align}
\bra{\mathcal{G}}c_{\textbf{k},\alpha}(t) c_{\textbf{l},\alpha}^\dagger(0)\ket{\mathcal{G}}=\frac{1}{2}e^{-iE(\textbf{k})t}\bigl(1+\cos\Theta(\textbf{k})\bigr).
\end{align}
Using this in \ref{D8}, and performing the sum over $\textbf{l}$, we get
\begin{align}
\bra{\mathcal{G}}c_{\textbf{r}_1,\alpha}(t) c_{\textbf{r}_2,\alpha}^\dagger(0)\ket{\mathcal{G}}=&\frac{1}{2N_xN_y}\sum_{\textbf{k}}e^{i\textbf{k}\cdot(\textbf{r}_1-\textbf{r}_2)}\notag\\
	&\cross e^{-iE(\textbf{k})t}\bigl(1+\cos\Theta(\textbf{k})\bigr).
\end{align}
Similarly, we obtain
\begin{align}
\bra{\mathcal{G}} c_{\textbf{r}_2,\alpha}^\dagger(0)c_{\textbf{r}_1,\alpha}(t)\ket{\mathcal{G}}=&\frac{1}{2N_xN_y}\sum_{\textbf{k}}e^{i\textbf{k}\cdot(\textbf{r}_1-\textbf{r}_2)}\notag\\
	&\cross e^{iE(\textbf{k})t}\bigl(1-\cos\Theta(\textbf{k})\bigr).
\end{align}
Plugging everything together, the retarded Green's function becomes
\begin{align}
G_{\alpha \alpha}^R(\textbf{r}_1,t;\textbf{r}_2,0)
	=-i\frac{\theta(t)}{2N_xN_y}\sum_{\textbf{k}}e^{i\textbf{k}\cdot(\textbf{r}_1-\textbf{r}_2)}\notag\\
	\cross\Bigl[e^{-iE(\textbf{k})t}\bigl(1+\cos\Theta(\textbf{k})\bigr)+e^{iE(\textbf{k})t}\bigl(1-\cos\Theta(\textbf{k})\bigr)\Bigr].
\end{align}

We can again take Fourier transform of this in $\textbf{r}_1-\textbf{r}_2$ and $t$. For the Fourier transform in time, we have to shift the frequency $\omega$ to complex frequency $\omega+i\epsilon$, where $\epsilon$ is a small positive number, and in the end we take the limit $\epsilon \rightarrow 0$. After doing this, we get
\begin{align}
&G_{\alpha\alpha}^R(\textbf{k},\omega)\notag\\
&=\frac{1}{2}\Bigl[\bigl(1+\cos\Theta(\textbf{k})\bigr)\Bigl(\mathcal{P}\Big(\frac{1}{\omega -E(\textbf{k})}\Bigr)- i\pi\delta(\omega -E(\textbf{k}))\Bigr)\notag\\
	&+\bigl(1-\cos\Theta(\textbf{k})\bigr)\Bigl(\mathcal{P}\Big(\frac{1}{\omega +E(\textbf{k})}\Bigr)- i\pi\delta(\omega +E(\textbf{k}))\Bigr)\Bigr],
\end{align}
where, we have again applied the identity in Eq. \ref{CPV}. Finally, we get the SPSF as
\begin{align}
A_{\alpha\alpha}(\textbf{k},\omega)&=-2\text{Im}G_{\alpha\alpha}^R(\textbf{k},\omega)\notag\\
&=\pi\bigl(1+\cos\Theta(\textbf{k})\bigr)\delta(\omega -E(\textbf{k}))\notag\\
&+\pi\bigl(1-\cos\Theta(\textbf{k})\bigr)\delta(\omega +E(\textbf{k})).
\end{align}
We can easily check that the above formula also satisfies the sum rule in Eq. \ref{sumrule}.
\section{SPSF for disordered Kitaev chain}\label{Disorder}

\begin{figure}[h]
  \centering
  {\includegraphics[width=\linewidth]{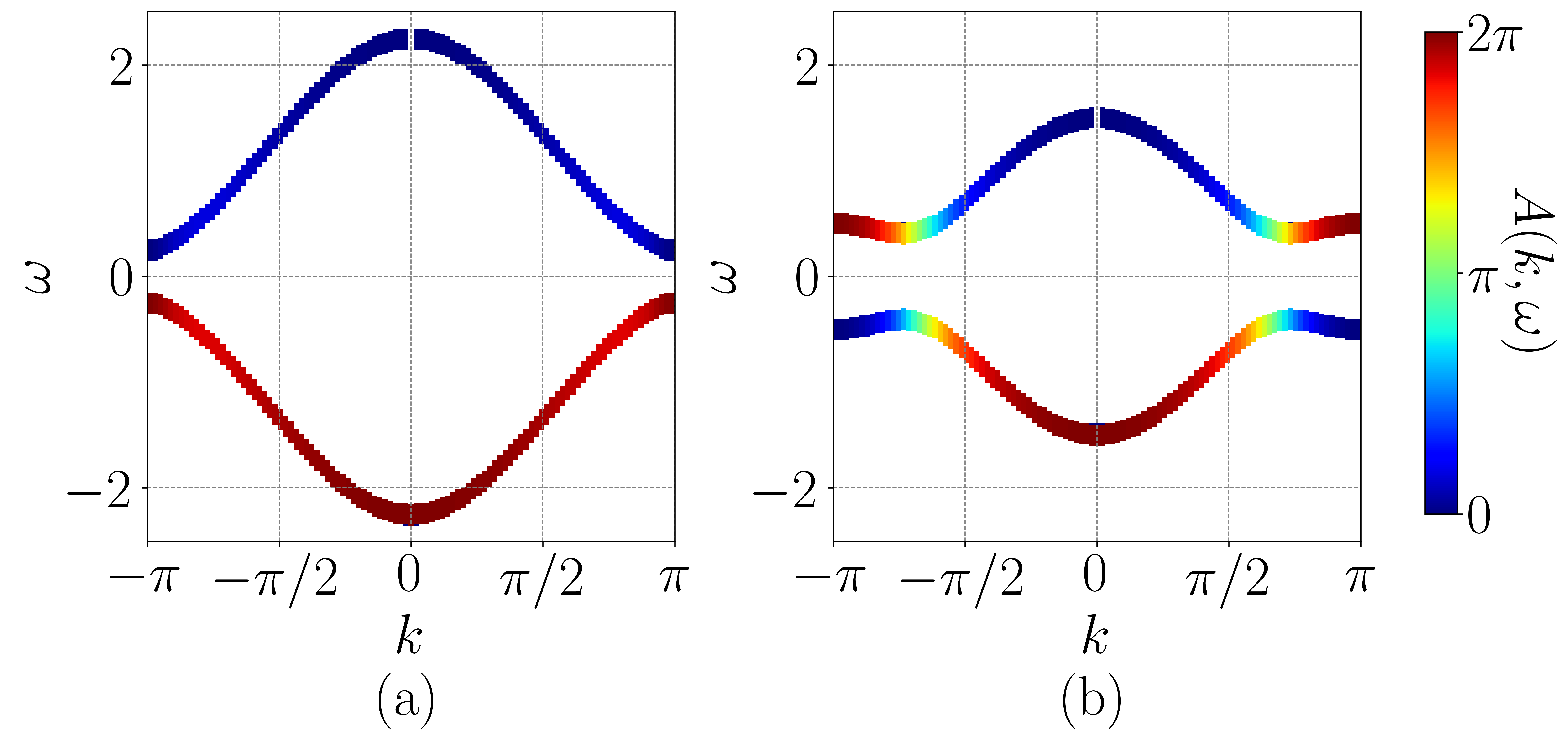}}
  \caption{Numerically computed SPSF  ($A(k,\omega)$) with $k,\omega$ for an ordered Kitaev chain of $N=101$ sites in (a) trivial phase $(\mu=2.5, J=1, \Delta=0.5)$, and (b) non-trivial phase $(\mu=1, J=1, \Delta=0.5)$ showing absence and presence of $\pi$ crossings, respectively.}
  \label{KD0P}
\end{figure}


\begin{figure}[h]
  \centering
  {\includegraphics[width=\linewidth]{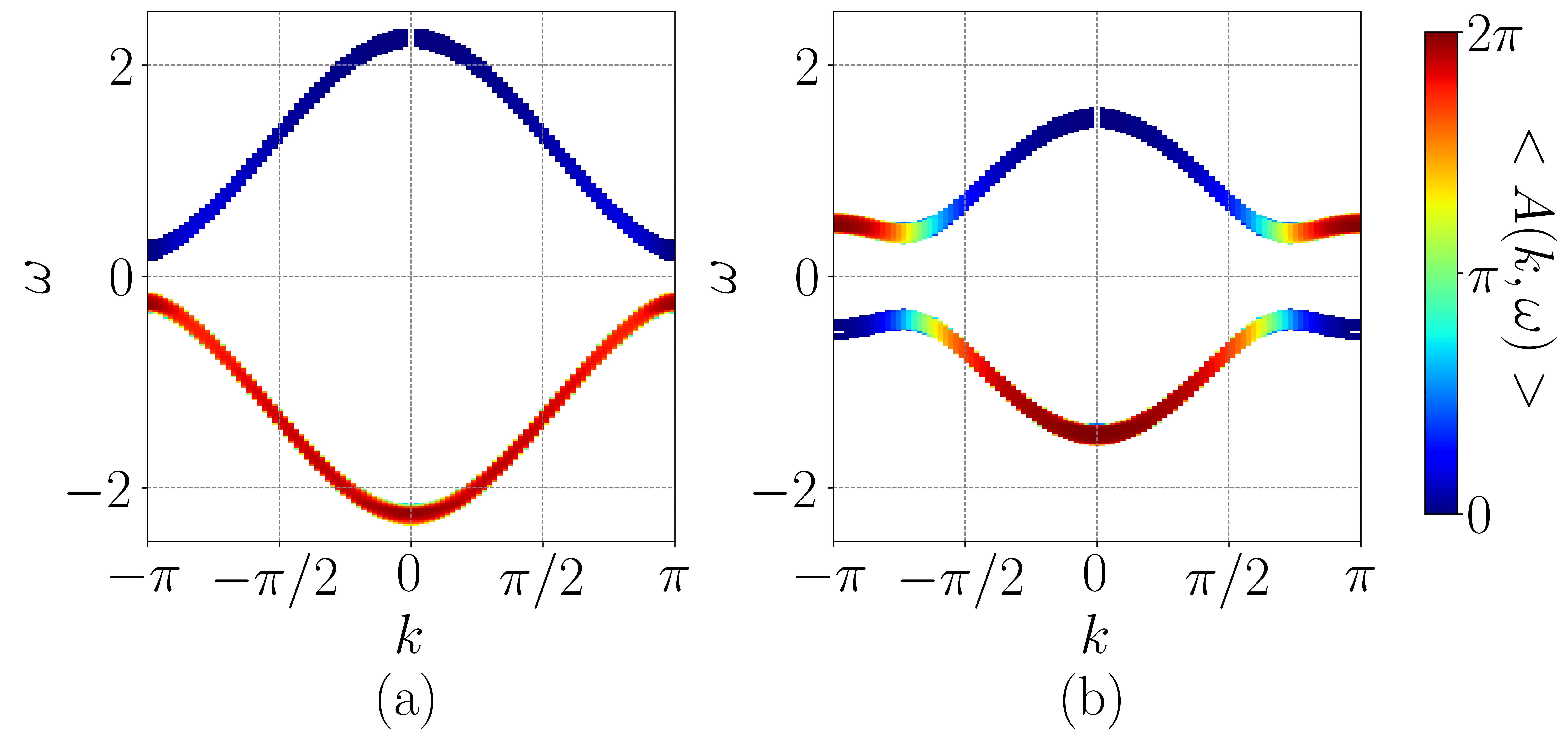}}
  \caption{Disorder averaged SPSF ($\langle A(k,\omega) \rangle$) with $k,\omega$ for a disordered Kitaev chain of $N=101$ sites in (a) trivial phase with $\langle \mu \rangle =2.5, J=1; \Delta=0.5$, and (b) non-trivial phase $\langle \mu \rangle=1,J=1; \Delta=0.5$. The relative standard deviation of disorder in $\mu$ is $20\%$. An averaging over 20 realizations of disorder is performed.}
  \label{KD10P}
\end{figure}

\begin{figure}[htbp]
  \centering
 {\includegraphics[width=\linewidth]{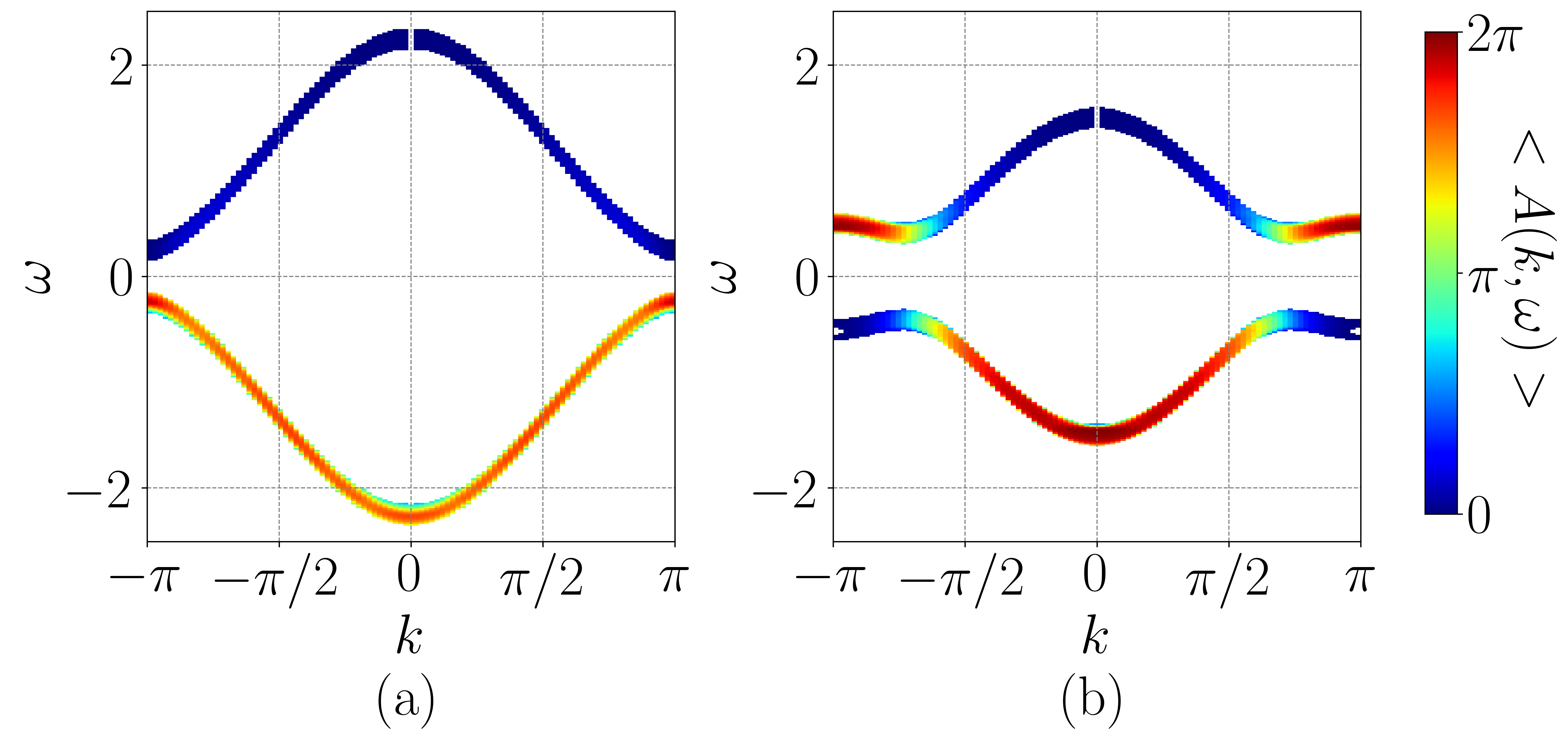}}
  \caption{Disorder averaged SPSF ($\langle A(k,\omega) \rangle$) with $k,\omega$ for a disordered Kitaev chain of $N=101$ sites in (a) trivial phase with $\langle \mu \rangle =2.5, J=1, \Delta=0.5$, and (b) non-trivial phase $\langle \mu \rangle=1, J=1, \Delta=0.5$. The relative standard deviation of disorder in $\mu$ is $40\%$. An averaging over 20 realizations of disorder is performed. }
  \label{KD40P}
\end{figure}

In this appendix, we calculate the SPSF in a disordered Kitaev chain, and compare it with an ordered case to check the validity of our proposed scheme in realistic materials with imperfections. For this, we numerically compute the SPSF for the Kitaev chain with some amount of randomness in chemical potential ($\mu$) or hopping amplitude ($J$) over the lattice sites. Randomness in hopping amplitudes or the chemical potential breaks the translation symmetry. The momentum quantum number $k$ is no longer an appropriate description in the absence of translation symmetry due to disorder. We quantify the disorder in any parameter by its mean and relative standard deviation, which is a ratio of the standard deviation and the mean. In our numerics, we first compute the retarded Green's function for all possible separations $x_1-x_2$ of a periodic Kitaev chain of fixed length and disorder realization. For multiple values of retarded Green's function at a fixed separation $x_1-x_2$, we take their average value. Naturally, there are more such combinations for small $x_1-x_2$, which results in better precision in the computed SPSF at large $k$. Since the time translation symmetry is still there, we can analytically take the Fourier transform in time as described in App. \ref{AppendixB}. We numerically find the Fourier transform of the retarded Green's function in space to obtain the SPSF for each disorder realization, and then perform another averaging over different disorder realizations. We denote the Fourier mode of space separation by index $k$. Fig.~\ref{KD0P} shows that the profiles of numerically computed $A(k,\omega)$ with $k,\omega$ for an ordered Kitaev chain in the trivial and non-trivial phase are similar to the analytical results in Fig. \ref{fig.3}.

In Figs.~\ref{KD10P} and \ref{KD40P}, we depict the disorder averaged SPSF ($\langle A(k,\omega) \rangle$) of the disordered Kitaev chain in the trivial and non-trivial phase for different disorder strength in $\mu$. Comparing the plots in Figs.~\ref{KD10P} and \ref{KD40P} to Fig.~\ref{KD0P}, we observe that the SPSF remains qualitatively the same to the ordered case even for a substantial amount of disorder. 
The signature of bulk topology in terms of the $\pi$ crossings of the value of $\langle A(k,\omega) \rangle$ remains mostly robust against disorder in the chemical potential as well as hopping parameters, the latter is not shown here.\\

\section{Haldane's Graphene model}\label{Haldane}
In this appendix, we study the SPSF for another 2D model of a topological insulator, namely, the Haldane model on the honeycomb lattice. Haldane proposed it to mimic the integer quantum Hall effect without a magnetic field. The $k$-space Hamiltonian of this model is given by \cite{Bernevig}
\begin{align}
H_G&=\sum_\textbf{k}\Psi_\textbf{k}^\dagger H_G(\textbf{k}) \Psi_\textbf{k},~{\rm where}\\
H_G(\textbf{k})&=\epsilon(\textbf{k})+d_i(\textbf{k})\sigma_i.
\end{align}
Here, $\Psi_k=\begin{pmatrix} c_{\textbf{k}\alpha}& c_{\textbf{k}\beta}\end{pmatrix}^T$, and
\begin{align}
\epsilon(\textbf{k})&=2t_2(\cos(\phi))(\cos(\textbf{k}\cdot\textbf{a}_1)+\cos(\textbf{k}\cdot\textbf{a}_2)\notag\\
&+\cos(\textbf{k}\cdot(\textbf{a}_1-\textbf{a}_2))),\notag\\
d_1(\textbf{k})&=\cos(\textbf{k}\cdot\textbf{a}_1)+\cos(\textbf{k}\cdot\textbf{a}_2)+1,\notag\\
d_2(\textbf{k})&=\sin(\textbf{k}\cdot\textbf{a}_1)+\sin(\textbf{k}\cdot\textbf{a}_2),\notag\\
d_3(\textbf{k})&=M+2t_2(\sin(\phi))(\sin(\textbf{k}\cdot\textbf{a}_1)-\sin(\textbf{k}\cdot\textbf{a}_2)\notag\\
&-\sin(\textbf{k}\cdot(\textbf{a}_1-\textbf{a}_2))).
\end{align}
Further, $\textbf{a}_1$ and $\textbf{a}_2$ are the lattice translation vectors of the honeycomb lattice. $t_1$ and $t_2$ are the nearest-neighbour and next-nearest-neighbour-hopping amplitudes, respectively. $M$ is the onsite energy, which has opposite signs for $\alpha$ and $\beta$ sublattice sites. The magnetic phase $\phi$ with the next-nearest-neighbor hopping $t_2$ breaks time-reversal symmetry without a net flux per plaquette, keeping the translational symmetry of the lattice. This model has non-trivial topological phases with a phase diagram similar to Fig.~\ref{ESSH_phasediagram} \cite{Bernevig}.

\begin{figure}[h]
  \centering
 {\includegraphics[width=\linewidth]{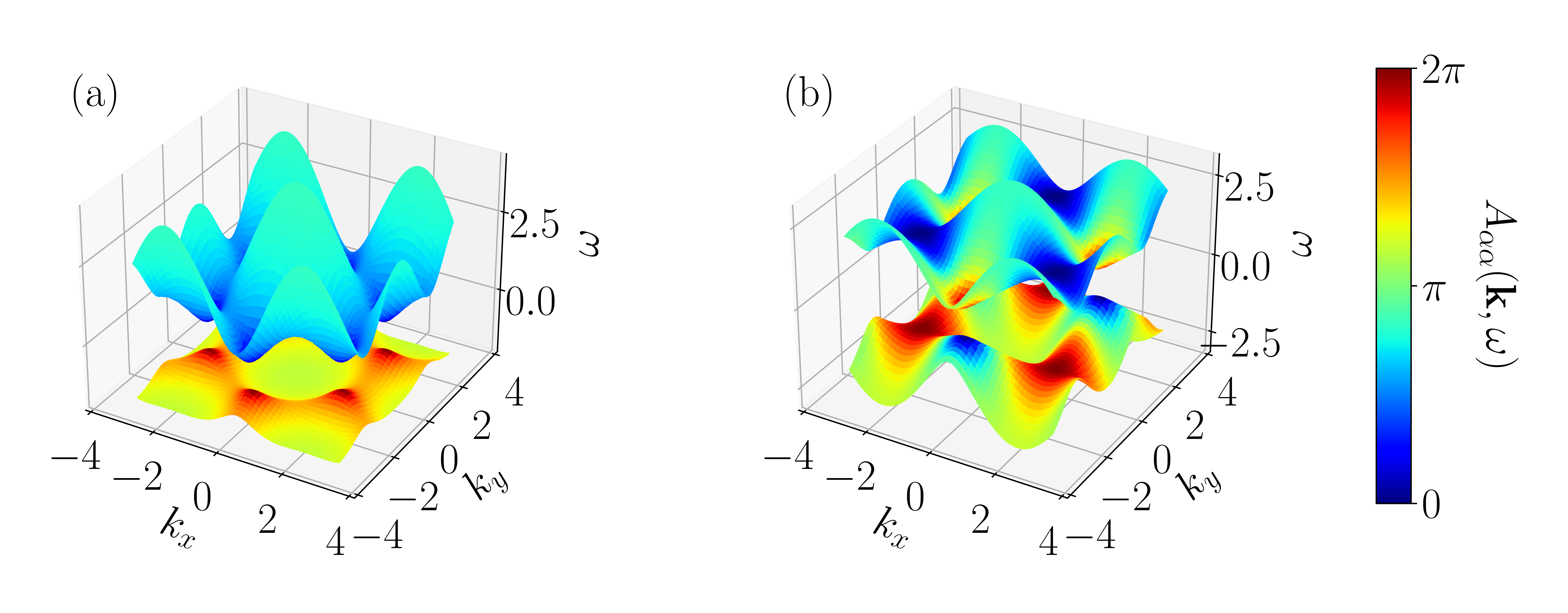}}
  \caption{Diagonal SPSF ($A_{\alpha \alpha}(\textbf{k},\omega)$) of the Haldane's graphene model with $\textbf{k}$, $\omega$ for (a) trivial phase ($\phi=0$) and (b) non-trivial phase ($\phi=\pi/2$). The parameters $t_2=1/\sqrt{27}$ and $M=0.5$ are same for both the plots.}
  \label{HaldaneSPSF}
\end{figure}
We calculate the diagonal SPSF, $A_{\alpha \alpha}(\textbf{k},\omega)$, numerically and see its behaviour in the trivial and non-trivial phases. The results are shown in Fig.~\ref{HaldaneSPSF}. We observe a similar trend in the behavior of the value of diagonal SPSF for the Haldane model as was observed in other models. In Fig.~\ref{HaldaneSPSF}(a), the SPSF in the trivial phase has values less than $\pi$ and greater than $\pi$ for upper and lower energy bands, respectively. On the other hand, for the non-trivial phase in Fig. \ref{HaldaneSPSF}(b), the SPSF takes values from $0$ to $2\pi$ in each energy band. There is this qualitative difference in the behavior of the SPSF in the trivial and non-trivial phases. This difference in behavior persists even if the parameters of the Hamiltonian are changed while staying in the same phases. It indicates that the signature of topology can be detected from the SPSF for different 2D topological matters with a topological phase transition separating the trivial and non-trivial phases.

\section{Physical realization of Kitaev chain: A four-band model}\label{Fourband}
All the models discussed so far consist of two energy bands. Here, we discuss the validity of our proposal for a topological matter with more than two bands. The Kitaev chain discussed in \ref{Kitaev} is a 1D topological superconductor that can be experimentally realized by interfacing a 2D topological insulator with a s-wave superconductor. Such a 1D hybrid structure has four energy bands, which, in the appropriate limit, can show the topological phases of the spinless Kitaev chain \cite{AliceaReview2012}.

\begin{figure}[h]
  \centering
 {\includegraphics[width=\linewidth]{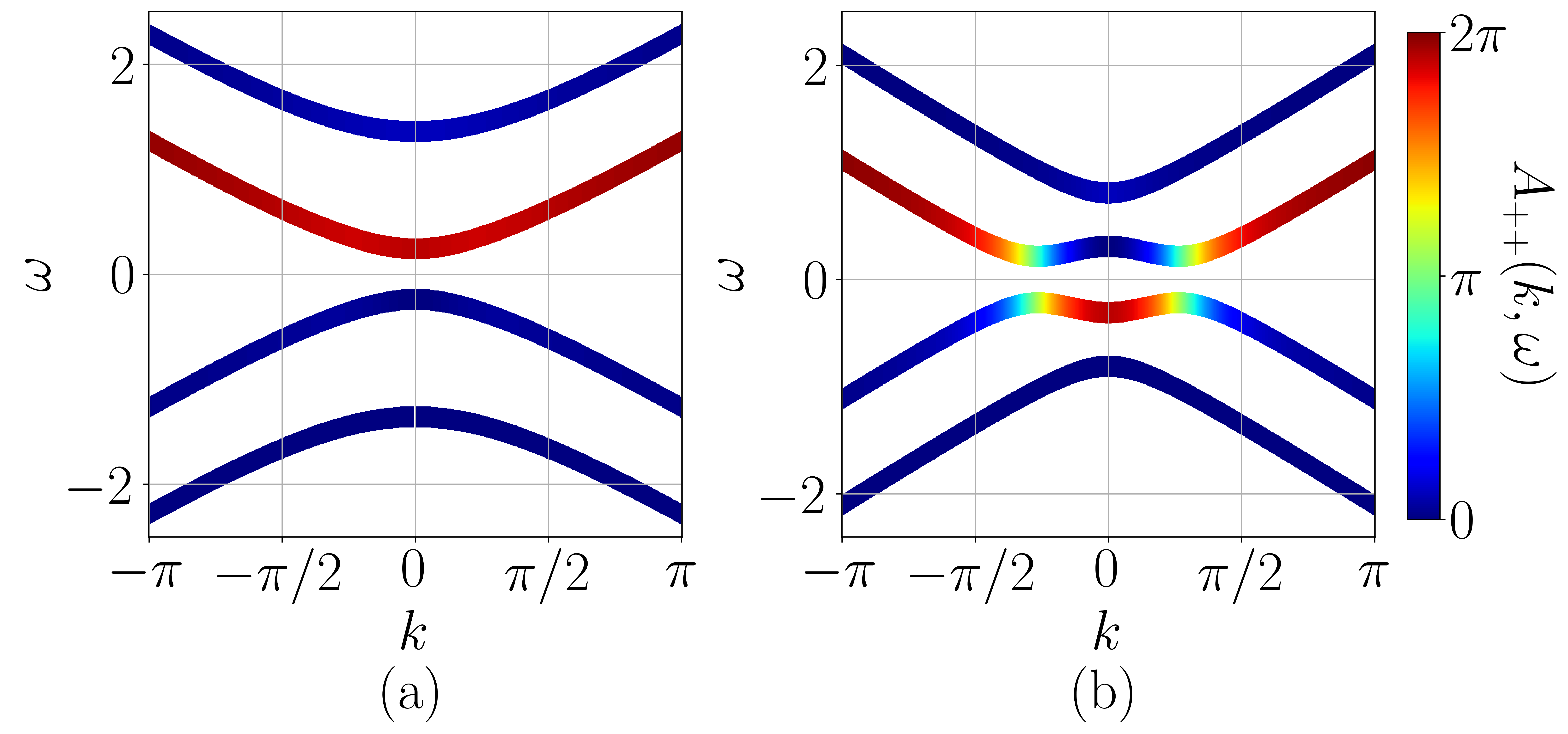}}
  \caption{Diagonal SPSF ($A_{++}(k,\omega)$) of a physical realization of Kitaev chain with $k$, $\omega$ for (a) trivial phase ($h=1.6$) and (b) non-trivial phase ($h=0.5$). The parameters $\mu=1$, $\Delta=0.5$, $v=1$ are same for both the plots.}
  \label{ExptKitaev}
\end{figure}

The $k$-space Hamiltonian of this model is given by 
\begin{align}
H_T&=\sum_k\Psi_k^\dagger H_T(k) \Psi_k, ~{\rm where}\\
H_T(k)&=\frac{1}{2}\begin{pmatrix}\epsilon_+(k)&0&\Delta_p(k)&\Delta_s(k)\\0&\epsilon_-(k)&-\Delta_s(k)&\Delta_p(k)\\\Delta_p(k)&-\Delta_s(k)&-\epsilon_+(k)&0\\\Delta_s(k)&\Delta_p(k)&0&-\epsilon_-(k)\end{pmatrix},
\end{align}
and $\Psi_k=\begin{pmatrix} \psi_+(k)& \psi_-(k)&\psi_+^\dagger(-k)&\psi_-^\dagger(-k)\end{pmatrix}^T$. Here, $\psi_\pm^\dagger(k)$ create fermions with energy $\epsilon_\pm(k)$, where $\epsilon_\pm(k)=-\mu\pm\sqrt{(vk)^2+h^2}$ are the energy of the edge modes of the 2D topological insulator in the absence of superconductor. Further, $\Delta_p(k)=vk\Delta/\sqrt{(vk)^2+h^2}$ and $\Delta_s(k)=h\Delta/\sqrt{(vk)^2+h^2}$ are proximity induced p-wave and s-wave pairing, where $\mu, v, h$ and $\Delta$ are, respectively, the chemical potential, the edge-state velocity of the 2D topological insulator, the Zeeman energy and the pairing of the s-wave superconductor. This hybrid structure generates a topological superconductor with four energy bands and supports a non-trivial topological phase. The model is in trivial phase for $h>\sqrt{\mu^2+\Delta^2}$ and in non-trivial phase for $h<\sqrt{\mu^2+\Delta^2}$ \cite{AliceaReview2012}. We compute the diagonal SPSF $A_{++}(k,\omega)$ for the half-filled ground state numerically. The results are presented in Fig.~\ref{ExptKitaev}.

We find that the diagonal SPSF $A_{++}(k,\omega)$ behaves qualitatively differently for the middle two bands of the model in the trivial and non-trivial phases. Let us denote the bands from 1 to 4 with increasing energies. In trivial phase $(h>\sqrt{\mu^2+\Delta^2})$, the value of SPSF for bands 2 and 3 are always less than $\pi$ and larger than $\pi$, respectively. On the other hand, in non-trivial phase $(h<\sqrt{\mu^2+\Delta^2})$, the behavior is qualitatively different for these two middle bands. The value of SPSF varies from $0$ to $2\pi$ by crossing $\pi$ twice for both the bands as $k$ goes from $-\pi$ to $\pi$. Such difference in behavior continues to be true for any other parameter values as long as $h>\sqrt{\mu^2+\Delta^2}$ for the trivial phase and $h<\sqrt{\mu^2+\Delta^2}$ for the non-trivial phase.

 \bibliography{references2}

\begin{thebibliography}{45}%
\makeatletter
\providecommand \@ifxundefined [1]{%
 \@ifx{#1\undefined}
}%
\providecommand \@ifnum [1]{%
 \ifnum #1\expandafter \@firstoftwo
 \else \expandafter \@secondoftwo
 \fi
}%
\providecommand \@ifx [1]{%
 \ifx #1\expandafter \@firstoftwo
 \else \expandafter \@secondoftwo
 \fi
}%
\providecommand \natexlab [1]{#1}%
\providecommand \enquote  [1]{``#1''}%
\providecommand \bibnamefont  [1]{#1}%
\providecommand \bibfnamefont [1]{#1}%
\providecommand \citenamefont [1]{#1}%
\providecommand \href@noop [0]{\@secondoftwo}%
\providecommand \href [0]{\begingroup \@sanitize@url \@href}%
\providecommand \@href[1]{\@@startlink{#1}\@@href}%
\providecommand \@@href[1]{\endgroup#1\@@endlink}%
\providecommand \@sanitize@url [0]{\catcode `\\12\catcode `\$12\catcode
  `\&12\catcode `\#12\catcode `\^12\catcode `\_12\catcode `\%12\relax}%
\providecommand \@@startlink[1]{}%
\providecommand \@@endlink[0]{}%
\providecommand \url  [0]{\begingroup\@sanitize@url \@url }%
\providecommand \@url [1]{\endgroup\@href {#1}{\urlprefix }}%
\providecommand \urlprefix  [0]{URL }%
\providecommand \Eprint [0]{\href }%
\providecommand \doibase [0]{https://doi.org/}%
\providecommand \selectlanguage [0]{\@gobble}%
\providecommand \bibinfo  [0]{\@secondoftwo}%
\providecommand \bibfield  [0]{\@secondoftwo}%
\providecommand \translation [1]{[#1]}%
\providecommand \BibitemOpen [0]{}%
\providecommand \bibitemStop [0]{}%
\providecommand \bibitemNoStop [0]{.\EOS\space}%
\providecommand \EOS [0]{\spacefactor3000\relax}%
\providecommand \BibitemShut  [1]{\csname bibitem#1\endcsname}%
\let\auto@bib@innerbib\@empty
\bibitem [{\citenamefont {Thouless}(1998)}]{thouless1998topological}%
  \BibitemOpen
  \bibfield  {author} {\bibinfo {author} {\bibfnamefont {D.}~\bibnamefont
  {Thouless}},\ }\href {https://books.google.co.in/books?id=6XC\_PBEXnAEC}
  {\emph {\bibinfo {title} {Topological Quantum Numbers in Nonrelativistic
  Physics}}}\ (\bibinfo  {publisher} {World Scientific},\ \bibinfo {year}
  {1998})\BibitemShut {NoStop}%
\bibitem [{\citenamefont {v.~Klitzing}\ \emph {et~al.}(1980)\citenamefont
  {v.~Klitzing}, \citenamefont {Dorda},\ and\ \citenamefont {Pepper}}]{QH}%
  \BibitemOpen
  \bibfield  {author} {\bibinfo {author} {\bibfnamefont {K.}~\bibnamefont
  {v.~Klitzing}}, \bibinfo {author} {\bibfnamefont {G.}~\bibnamefont {Dorda}},\
  and\ \bibinfo {author} {\bibfnamefont {M.}~\bibnamefont {Pepper}},\
  }\bibfield  {title} {\bibinfo {title} {New method for high-accuracy
  determination of the fine-structure constant based on quantized {Hall}
  resistance},\ }\href {https://doi.org/10.1103/PhysRevLett.45.494} {\bibfield
  {journal} {\bibinfo  {journal} {Phys. Rev. Lett.}\ }\textbf {\bibinfo
  {volume} {45}},\ \bibinfo {pages} {494} (\bibinfo {year} {1980})}\BibitemShut
  {NoStop}%
\bibitem [{\citenamefont {Thouless}\ \emph {et~al.}(1982)\citenamefont
  {Thouless}, \citenamefont {Kohmoto}, \citenamefont {Nightingale},\ and\
  \citenamefont {den Nijs}}]{TKNN}%
  \BibitemOpen
  \bibfield  {author} {\bibinfo {author} {\bibfnamefont {D.~J.}\ \bibnamefont
  {Thouless}}, \bibinfo {author} {\bibfnamefont {M.}~\bibnamefont {Kohmoto}},
  \bibinfo {author} {\bibfnamefont {M.~P.}\ \bibnamefont {Nightingale}},\ and\
  \bibinfo {author} {\bibfnamefont {M.}~\bibnamefont {den Nijs}},\ }\bibfield
  {title} {\bibinfo {title} {Quantized {Hall} conductance in a two-dimensional
  periodic potential},\ }\href {https://doi.org/10.1103/PhysRevLett.49.405}
  {\bibfield  {journal} {\bibinfo  {journal} {Phys. Rev. Lett.}\ }\textbf
  {\bibinfo {volume} {49}},\ \bibinfo {pages} {405} (\bibinfo {year}
  {1982})}\BibitemShut {NoStop}%
\bibitem [{\citenamefont {Asbóth}\ \emph {et~al.}(2016)\citenamefont
  {Asbóth}, \citenamefont {Oroszlány},\ and\ \citenamefont
  {Pályi}}]{Asboth}%
  \BibitemOpen
  \bibfield  {author} {\bibinfo {author} {\bibfnamefont {J.~K.}\ \bibnamefont
  {Asbóth}}, \bibinfo {author} {\bibfnamefont {L.}~\bibnamefont
  {Oroszlány}},\ and\ \bibinfo {author} {\bibfnamefont {A.}~\bibnamefont
  {Pályi}},\ }\href {https://doi.org/10.1007/978-3-319-25607-8} {\emph
  {\bibinfo {title} {A Short Course on Topological Insulators}}}\ (\bibinfo
  {publisher} {Springer},\ \bibinfo {year} {2016})\BibitemShut {NoStop}%
\bibitem [{\citenamefont {King-Smith}\ and\ \citenamefont
  {Vanderbilt}(1993)}]{KingSmith1993}%
  \BibitemOpen
  \bibfield  {author} {\bibinfo {author} {\bibfnamefont {R.~D.}\ \bibnamefont
  {King-Smith}}\ and\ \bibinfo {author} {\bibfnamefont {D.}~\bibnamefont
  {Vanderbilt}},\ }\bibfield  {title} {\bibinfo {title} {Theory of polarization
  of crystalline solids},\ }\href {https://doi.org/10.1103/PhysRevB.47.1651}
  {\bibfield  {journal} {\bibinfo  {journal} {Phys. Rev. B}\ }\textbf {\bibinfo
  {volume} {47}},\ \bibinfo {pages} {1651} (\bibinfo {year}
  {1993})}\BibitemShut {NoStop}%
\bibitem [{\citenamefont {Thouless}(1983)}]{Thouless1983}%
  \BibitemOpen
  \bibfield  {author} {\bibinfo {author} {\bibfnamefont {D.~J.}\ \bibnamefont
  {Thouless}},\ }\bibfield  {title} {\bibinfo {title} {Quantization of particle
  transport},\ }\href {https://doi.org/10.1103/PhysRevB.27.6083} {\bibfield
  {journal} {\bibinfo  {journal} {Phys. Rev. B}\ }\textbf {\bibinfo {volume}
  {27}},\ \bibinfo {pages} {6083} (\bibinfo {year} {1983})}\BibitemShut
  {NoStop}%
\bibitem [{\citenamefont {Xiao}\ \emph {et~al.}(2014)\citenamefont {Xiao},
  \citenamefont {Zhang},\ and\ \citenamefont {Chan}}]{Chan2014}%
  \BibitemOpen
  \bibfield  {author} {\bibinfo {author} {\bibfnamefont {M.}~\bibnamefont
  {Xiao}}, \bibinfo {author} {\bibfnamefont {Z.~Q.}\ \bibnamefont {Zhang}},\
  and\ \bibinfo {author} {\bibfnamefont {C.~T.}\ \bibnamefont {Chan}},\
  }\bibfield  {title} {\bibinfo {title} {Surface impedance and bulk band
  geometric phases in one-dimensional systems},\ }\href
  {https://doi.org/10.1103/PhysRevX.4.021017} {\bibfield  {journal} {\bibinfo
  {journal} {Phys. Rev. X}\ }\textbf {\bibinfo {volume} {4}},\ \bibinfo {pages}
  {021017} (\bibinfo {year} {2014})}\BibitemShut {NoStop}%
\bibitem [{\citenamefont {Gao}\ \emph {et~al.}(2015)\citenamefont {Gao},
  \citenamefont {Xiao}, \citenamefont {Chan},\ and\ \citenamefont
  {Tam}}]{Chan2015}%
  \BibitemOpen
  \bibfield  {author} {\bibinfo {author} {\bibfnamefont {W.~S.}\ \bibnamefont
  {Gao}}, \bibinfo {author} {\bibfnamefont {M.}~\bibnamefont {Xiao}}, \bibinfo
  {author} {\bibfnamefont {C.~T.}\ \bibnamefont {Chan}},\ and\ \bibinfo
  {author} {\bibfnamefont {W.~Y.}\ \bibnamefont {Tam}},\ }\bibfield  {title}
  {\bibinfo {title} {Determination of {Zak} phase by reflection phase in {1D}
  photonic crystals},\ }\href {https://doi.org/10.1364/OL.40.005259} {\bibfield
   {journal} {\bibinfo  {journal} {Opt. Lett.}\ }\textbf {\bibinfo {volume}
  {40}},\ \bibinfo {pages} {5259} (\bibinfo {year} {2015})}\BibitemShut
  {NoStop}%
\bibitem [{\citenamefont {Kitaev}(2001)}]{Kitaev}%
  \BibitemOpen
  \bibfield  {author} {\bibinfo {author} {\bibfnamefont {A.~Y.}\ \bibnamefont
  {Kitaev}},\ }\bibfield  {title} {\bibinfo {title} {Unpaired {Majorana}
  fermions in quantum wires},\ }\href
  {https://doi.org/10.1070/1063-7869/44/10S/S29} {\bibfield  {journal}
  {\bibinfo  {journal} {Phys.-Usp}\ }\textbf {\bibinfo {volume} {44}},\
  \bibinfo {pages} {131} (\bibinfo {year} {2001})}\BibitemShut {NoStop}%
\bibitem [{\citenamefont {Bolech}\ and\ \citenamefont
  {Demler}(2007)}]{BolechPRL2007}%
  \BibitemOpen
  \bibfield  {author} {\bibinfo {author} {\bibfnamefont {C.~J.}\ \bibnamefont
  {Bolech}}\ and\ \bibinfo {author} {\bibfnamefont {E.}~\bibnamefont
  {Demler}},\ }\bibfield  {title} {\bibinfo {title} {Observing {Majorana} bound
  states in $p$-wave superconductors using noise measurements in tunneling
  experiments},\ }\href {https://doi.org/10.1103/PhysRevLett.98.237002}
  {\bibfield  {journal} {\bibinfo  {journal} {Phys. Rev. Lett.}\ }\textbf
  {\bibinfo {volume} {98}},\ \bibinfo {pages} {237002} (\bibinfo {year}
  {2007})}\BibitemShut {NoStop}%
\bibitem [{\citenamefont {Law}\ \emph {et~al.}(2009)\citenamefont {Law},
  \citenamefont {Lee},\ and\ \citenamefont {Ng}}]{LawPRL2009}%
  \BibitemOpen
  \bibfield  {author} {\bibinfo {author} {\bibfnamefont {K.~T.}\ \bibnamefont
  {Law}}, \bibinfo {author} {\bibfnamefont {P.~A.}\ \bibnamefont {Lee}},\ and\
  \bibinfo {author} {\bibfnamefont {T.~K.}\ \bibnamefont {Ng}},\ }\bibfield
  {title} {\bibinfo {title} {Majorana fermion induced resonant {Andreev}
  reflection},\ }\href {https://doi.org/10.1103/PhysRevLett.103.237001}
  {\bibfield  {journal} {\bibinfo  {journal} {Phys. Rev. Lett.}\ }\textbf
  {\bibinfo {volume} {103}},\ \bibinfo {pages} {237001} (\bibinfo {year}
  {2009})}\BibitemShut {NoStop}%
\bibitem [{\citenamefont {Roy}\ \emph {et~al.}(2012)\citenamefont {Roy},
  \citenamefont {Bolech},\ and\ \citenamefont {Shah}}]{RoyPRB2012}%
  \BibitemOpen
  \bibfield  {author} {\bibinfo {author} {\bibfnamefont {D.}~\bibnamefont
  {Roy}}, \bibinfo {author} {\bibfnamefont {C.~J.}\ \bibnamefont {Bolech}},\
  and\ \bibinfo {author} {\bibfnamefont {N.}~\bibnamefont {Shah}},\ }\bibfield
  {title} {\bibinfo {title} {Majorana fermions in a topological superconducting
  wire out of equilibrium: Exact microscopic transport analysis of a $p$-wave
  open chain coupled to normal leads},\ }\href
  {https://doi.org/10.1103/PhysRevB.86.094503} {\bibfield  {journal} {\bibinfo
  {journal} {Phys. Rev. B}\ }\textbf {\bibinfo {volume} {86}},\ \bibinfo
  {pages} {094503} (\bibinfo {year} {2012})}\BibitemShut {NoStop}%
\bibitem [{\citenamefont {Kells}\ \emph {et~al.}(2012)\citenamefont {Kells},
  \citenamefont {Meidan},\ and\ \citenamefont {Brouwer}}]{Kells2012}%
  \BibitemOpen
  \bibfield  {author} {\bibinfo {author} {\bibfnamefont {G.}~\bibnamefont
  {Kells}}, \bibinfo {author} {\bibfnamefont {D.}~\bibnamefont {Meidan}},\ and\
  \bibinfo {author} {\bibfnamefont {P.~W.}\ \bibnamefont {Brouwer}},\
  }\bibfield  {title} {\bibinfo {title} {Near-zero-energy end states in
  topologically trivial spin-orbit coupled superconducting nanowires with a
  smooth confinement},\ }\href {https://doi.org/10.1103/PhysRevB.86.100503}
  {\bibfield  {journal} {\bibinfo  {journal} {Phys. Rev. B}\ }\textbf {\bibinfo
  {volume} {86}},\ \bibinfo {pages} {100503} (\bibinfo {year}
  {2012})}\BibitemShut {NoStop}%
\bibitem [{\citenamefont {Roy}\ \emph {et~al.}(2013)\citenamefont {Roy},
  \citenamefont {Bondyopadhaya},\ and\ \citenamefont {Tewari}}]{Roy2013}%
  \BibitemOpen
  \bibfield  {author} {\bibinfo {author} {\bibfnamefont {D.}~\bibnamefont
  {Roy}}, \bibinfo {author} {\bibfnamefont {N.}~\bibnamefont {Bondyopadhaya}},\
  and\ \bibinfo {author} {\bibfnamefont {S.}~\bibnamefont {Tewari}},\
  }\bibfield  {title} {\bibinfo {title} {Topologically trivial zero-bias
  conductance peak in semiconductor {Majorana} wires from boundary effects},\
  }\href {https://doi.org/10.1103/PhysRevB.88.020502} {\bibfield  {journal}
  {\bibinfo  {journal} {Phys. Rev. B}\ }\textbf {\bibinfo {volume} {88}},\
  \bibinfo {pages} {020502(R)} (\bibinfo {year} {2013})}\BibitemShut {NoStop}%
\bibitem [{\citenamefont {Bondyopadhaya}\ and\ \citenamefont
  {Roy}(2019)}]{Bondyopadhaya2019}%
  \BibitemOpen
  \bibfield  {author} {\bibinfo {author} {\bibfnamefont {N.}~\bibnamefont
  {Bondyopadhaya}}\ and\ \bibinfo {author} {\bibfnamefont {D.}~\bibnamefont
  {Roy}},\ }\bibfield  {title} {\bibinfo {title} {Dynamics of hybrid junctions
  of {Majorana} wires},\ }\href {https://doi.org/10.1103/PhysRevB.99.214514}
  {\bibfield  {journal} {\bibinfo  {journal} {Phys. Rev. B}\ }\textbf {\bibinfo
  {volume} {99}},\ \bibinfo {pages} {214514} (\bibinfo {year}
  {2019})}\BibitemShut {NoStop}%
\bibitem [{\citenamefont {Chen}\ \emph {et~al.}(2020)\citenamefont {Chen},
  \citenamefont {Gu}, \citenamefont {Li}, \citenamefont {Du},\ and\
  \citenamefont {Yang}}]{ARPES1}%
  \BibitemOpen
  \bibfield  {author} {\bibinfo {author} {\bibfnamefont {Y.}~\bibnamefont
  {Chen}}, \bibinfo {author} {\bibfnamefont {X.}~\bibnamefont {Gu}}, \bibinfo
  {author} {\bibfnamefont {Y.}~\bibnamefont {Li}}, \bibinfo {author}
  {\bibfnamefont {X.}~\bibnamefont {Du}},\ and\ \bibinfo {author}
  {\bibfnamefont {L.}~\bibnamefont {Yang}},\ }\bibfield  {title} {\bibinfo
  {title} {Recent advances in topological quantum materials by angle-resolved
  photoemission spectroscopy},\ }\href {https://doi.org/S2590238520303659}
  {\bibfield  {journal} {\bibinfo  {journal} {Matter}\ }\textbf {\bibinfo
  {volume} {3}},\ \bibinfo {pages} {1114} (\bibinfo {year} {2020})}\BibitemShut
  {NoStop}%
\bibitem [{\citenamefont {Lv}\ \emph {et~al.}(2019)\citenamefont {Lv},
  \citenamefont {Qian},\ and\ \citenamefont {Ding}}]{ARPES2}%
  \BibitemOpen
  \bibfield  {author} {\bibinfo {author} {\bibfnamefont {B.~Q.}\ \bibnamefont
  {Lv}}, \bibinfo {author} {\bibfnamefont {T.}~\bibnamefont {Qian}},\ and\
  \bibinfo {author} {\bibfnamefont {H.}~\bibnamefont {Ding}},\ }\bibfield
  {title} {\bibinfo {title} {Angle-resolved photoemission spectroscopy and its
  applications to topological materials},\ }\href
  {https://doi.org/s42254-019-0088-5} {\bibfield  {journal} {\bibinfo
  {journal} {Nat. Rev. Phys.}\ }\textbf {\bibinfo {volume} {1}},\ \bibinfo
  {pages} {609} (\bibinfo {year} {2019})}\BibitemShut {NoStop}%
\bibitem [{\citenamefont {Sobota}\ \emph {et~al.}(2021)\citenamefont {Sobota},
  \citenamefont {He},\ and\ \citenamefont {Shen}}]{Sobota2021}%
  \BibitemOpen
  \bibfield  {author} {\bibinfo {author} {\bibfnamefont {J.~A.}\ \bibnamefont
  {Sobota}}, \bibinfo {author} {\bibfnamefont {Y.}~\bibnamefont {He}},\ and\
  \bibinfo {author} {\bibfnamefont {Z.-X.}\ \bibnamefont {Shen}},\ }\bibfield
  {title} {\bibinfo {title} {Angle-resolved photoemission studies of quantum
  materials},\ }\href {https://doi.org/10.1103/RevModPhys.93.025006} {\bibfield
   {journal} {\bibinfo  {journal} {Rev. Mod. Phys.}\ }\textbf {\bibinfo
  {volume} {93}},\ \bibinfo {pages} {025006} (\bibinfo {year}
  {2021})}\BibitemShut {NoStop}%
\bibitem [{\citenamefont {Buchs}\ \emph {et~al.}(2009)\citenamefont {Buchs},
  \citenamefont {Bercioux}, \citenamefont {Ruffieux}, \citenamefont
  {Gr\"oning}, \citenamefont {Grabert},\ and\ \citenamefont
  {Gr\"oning}}]{Buchs2009}%
  \BibitemOpen
  \bibfield  {author} {\bibinfo {author} {\bibfnamefont {G.}~\bibnamefont
  {Buchs}}, \bibinfo {author} {\bibfnamefont {D.}~\bibnamefont {Bercioux}},
  \bibinfo {author} {\bibfnamefont {P.}~\bibnamefont {Ruffieux}}, \bibinfo
  {author} {\bibfnamefont {P.}~\bibnamefont {Gr\"oning}}, \bibinfo {author}
  {\bibfnamefont {H.}~\bibnamefont {Grabert}},\ and\ \bibinfo {author}
  {\bibfnamefont {O.}~\bibnamefont {Gr\"oning}},\ }\bibfield  {title} {\bibinfo
  {title} {Electron scattering in intrananotube quantum dots},\ }\href
  {https://doi.org/10.1103/PhysRevLett.102.245505} {\bibfield  {journal}
  {\bibinfo  {journal} {Phys. Rev. Lett.}\ }\textbf {\bibinfo {volume} {102}},\
  \bibinfo {pages} {245505} (\bibinfo {year} {2009})}\BibitemShut {NoStop}%
\bibitem [{\citenamefont {Mikhail}\ \emph {et~al.}(2022)\citenamefont
  {Mikhail}, \citenamefont {Voisin}, \citenamefont {Medar}, \citenamefont
  {Buchs}, \citenamefont {Rogge},\ and\ \citenamefont
  {Rachel}}]{InteractingSSH}%
  \BibitemOpen
  \bibfield  {author} {\bibinfo {author} {\bibfnamefont {D.}~\bibnamefont
  {Mikhail}}, \bibinfo {author} {\bibfnamefont {B.}~\bibnamefont {Voisin}},
  \bibinfo {author} {\bibfnamefont {D.~D.~S.}\ \bibnamefont {Medar}}, \bibinfo
  {author} {\bibfnamefont {G.}~\bibnamefont {Buchs}}, \bibinfo {author}
  {\bibfnamefont {S.}~\bibnamefont {Rogge}},\ and\ \bibinfo {author}
  {\bibfnamefont {S.}~\bibnamefont {Rachel}},\ }\bibfield  {title} {\bibinfo
  {title} {Quasiparticle excitations in one-dimensional interacting topological
  insulator: Application for dopant-based quantum simulation},\ }\href
  {https://doi.org/10.1103/PhysRevB.106.195408} {\bibfield  {journal} {\bibinfo
   {journal} {Phys. Rev. B}\ }\textbf {\bibinfo {volume} {106}},\ \bibinfo
  {pages} {195408} (\bibinfo {year} {2022})}\BibitemShut {NoStop}%
\bibitem [{\citenamefont {Xia}\ \emph {et~al.}(2009)\citenamefont {Xia},
  \citenamefont {Qian}, \citenamefont {Hseih}, \citenamefont {Wray},
  \citenamefont {Pal}, \citenamefont {Lin}, \citenamefont {Bansil},
  \citenamefont {Grauer}, \citenamefont {Hor}, \citenamefont {Cava},\ and\
  \citenamefont {Hasan}}]{Xia2009}%
  \BibitemOpen
  \bibfield  {author} {\bibinfo {author} {\bibfnamefont {Y.}~\bibnamefont
  {Xia}}, \bibinfo {author} {\bibfnamefont {D.}~\bibnamefont {Qian}}, \bibinfo
  {author} {\bibfnamefont {D.}~\bibnamefont {Hseih}}, \bibinfo {author}
  {\bibfnamefont {L.}~\bibnamefont {Wray}}, \bibinfo {author} {\bibfnamefont
  {A.}~\bibnamefont {Pal}}, \bibinfo {author} {\bibfnamefont {H.}~\bibnamefont
  {Lin}}, \bibinfo {author} {\bibfnamefont {A.}~\bibnamefont {Bansil}},
  \bibinfo {author} {\bibfnamefont {D.}~\bibnamefont {Grauer}}, \bibinfo
  {author} {\bibfnamefont {Y.~S.}\ \bibnamefont {Hor}}, \bibinfo {author}
  {\bibfnamefont {R.~J.}\ \bibnamefont {Cava}},\ and\ \bibinfo {author}
  {\bibfnamefont {M.~Z.}\ \bibnamefont {Hasan}},\ }\bibfield  {title} {\bibinfo
  {title} {Observation of a large-gap topological-insulator class with a single
  {Dirac} cone on the surface},\ }\href {https://doi.org/10.1038/nphys1274}
  {\bibfield  {journal} {\bibinfo  {journal} {Nat. Phys.}\ }\textbf {\bibinfo
  {volume} {5}},\ \bibinfo {pages} {398} (\bibinfo {year} {2009})}\BibitemShut
  {NoStop}%
\bibitem [{\citenamefont {Chen}\ \emph {et~al.}(2010)\citenamefont {Chen},
  \citenamefont {Liu},\ and\ \citenamefont {{\it et al.}}}]{Chen2010}%
  \BibitemOpen
  \bibfield  {author} {\bibinfo {author} {\bibfnamefont {Y.~L.}\ \bibnamefont
  {Chen}}, \bibinfo {author} {\bibfnamefont {Z.~K.}\ \bibnamefont {Liu}},\ and\
  \bibinfo {author} {\bibnamefont {{\it et al.}}},\ }\bibfield  {title}
  {\bibinfo {title} {Single {Dirac} cone topological surface state and unusual
  thermoelectric property of compounds from a new topological insulator
  family},\ }\href {https://doi.org/10.1103/PhysRevLett.105.266401} {\bibfield
  {journal} {\bibinfo  {journal} {Phys. Rev. Lett.}\ }\textbf {\bibinfo
  {volume} {105}},\ \bibinfo {pages} {266401} (\bibinfo {year}
  {2010})}\BibitemShut {NoStop}%
\bibitem [{\citenamefont {Xu}\ \emph {et~al.}(2015)\citenamefont {Xu},
  \citenamefont {Belopolski},\ and\ \citenamefont {{\it et al.}}}]{SuXu2015}%
  \BibitemOpen
  \bibfield  {author} {\bibinfo {author} {\bibfnamefont {S.-Y.}\ \bibnamefont
  {Xu}}, \bibinfo {author} {\bibfnamefont {I.}~\bibnamefont {Belopolski}},\
  and\ \bibinfo {author} {\bibnamefont {{\it et al.}}},\ }\bibfield  {title}
  {\bibinfo {title} {Discovery of a {Weyl} fermion semimetal and topological
  {Fermi} arcs},\ }\href {https://doi.org/10.1126/science.aaa9297} {\bibfield
  {journal} {\bibinfo  {journal} {Science}\ }\textbf {\bibinfo {volume}
  {349}},\ \bibinfo {pages} {613} (\bibinfo {year} {2015})}\BibitemShut
  {NoStop}%
\bibitem [{\citenamefont {Yin}\ \emph {et~al.}(2021)\citenamefont {Yin},
  \citenamefont {Pan},\ and\ \citenamefont {Hasan}}]{Yin21}%
  \BibitemOpen
  \bibfield  {author} {\bibinfo {author} {\bibfnamefont {J.~X.}\ \bibnamefont
  {Yin}}, \bibinfo {author} {\bibfnamefont {S.~H.}\ \bibnamefont {Pan}},\ and\
  \bibinfo {author} {\bibfnamefont {Z.~M.}\ \bibnamefont {Hasan}},\ }\bibfield
  {title} {\bibinfo {title} {Probing topological quantum matter with scanning
  tunnelling microscopy},\ }\href {https://doi.org/10.1038/s42254-021-00293-7}
  {\bibfield  {journal} {\bibinfo  {journal} {Nat. Rev. Phys.}\ }\textbf
  {\bibinfo {volume} {3}},\ \bibinfo {pages} {249} (\bibinfo {year}
  {2021})}\BibitemShut {NoStop}%
\bibitem [{\citenamefont {Roushan}\ \emph {et~al.}(2009)\citenamefont
  {Roushan}, \citenamefont {Seo}, \citenamefont {Parker}, \citenamefont {Hor},
  \citenamefont {Hsieh}, \citenamefont {Qian}, \citenamefont {Richardella},
  \citenamefont {Hasan}, \citenamefont {Cava},\ and\ \citenamefont
  {Yazdani}}]{Roushan2009}%
  \BibitemOpen
  \bibfield  {author} {\bibinfo {author} {\bibfnamefont {P.}~\bibnamefont
  {Roushan}}, \bibinfo {author} {\bibfnamefont {J.}~\bibnamefont {Seo}},
  \bibinfo {author} {\bibfnamefont {C.~V.}\ \bibnamefont {Parker}}, \bibinfo
  {author} {\bibfnamefont {Y.~S.}\ \bibnamefont {Hor}}, \bibinfo {author}
  {\bibfnamefont {D.}~\bibnamefont {Hsieh}}, \bibinfo {author} {\bibfnamefont
  {D.}~\bibnamefont {Qian}}, \bibinfo {author} {\bibfnamefont {A.}~\bibnamefont
  {Richardella}}, \bibinfo {author} {\bibfnamefont {M.~Z.}\ \bibnamefont
  {Hasan}}, \bibinfo {author} {\bibfnamefont {R.~J.}\ \bibnamefont {Cava}},\
  and\ \bibinfo {author} {\bibfnamefont {A.}~\bibnamefont {Yazdani}},\
  }\bibfield  {title} {\bibinfo {title} {Topological surface states protected
  from backscattering by chiral spin texture},\ }\href
  {https://doi.org/10.1038/nature08308} {\bibfield  {journal} {\bibinfo
  {journal} {Nature (London)}\ }\textbf {\bibinfo {volume} {460}},\ \bibinfo
  {pages} {1106} (\bibinfo {year} {2009})}\BibitemShut {NoStop}%
\bibitem [{\citenamefont {Alpichshev}\ \emph {et~al.}(2010)\citenamefont
  {Alpichshev}, \citenamefont {Analytis}, \citenamefont {Chu}, \citenamefont
  {Fisher}, \citenamefont {Chen}, \citenamefont {Shen}, \citenamefont {Fang},\
  and\ \citenamefont {Kapitulnik}}]{Alpichshev2010}%
  \BibitemOpen
  \bibfield  {author} {\bibinfo {author} {\bibfnamefont {Z.}~\bibnamefont
  {Alpichshev}}, \bibinfo {author} {\bibfnamefont {J.~G.}\ \bibnamefont
  {Analytis}}, \bibinfo {author} {\bibfnamefont {J.-H.}\ \bibnamefont {Chu}},
  \bibinfo {author} {\bibfnamefont {I.~R.}\ \bibnamefont {Fisher}}, \bibinfo
  {author} {\bibfnamefont {Y.~L.}\ \bibnamefont {Chen}}, \bibinfo {author}
  {\bibfnamefont {Z.~X.}\ \bibnamefont {Shen}}, \bibinfo {author}
  {\bibfnamefont {A.}~\bibnamefont {Fang}},\ and\ \bibinfo {author}
  {\bibfnamefont {A.}~\bibnamefont {Kapitulnik}},\ }\bibfield  {title}
  {\bibinfo {title} {Stm imaging of electronic waves on the surface of
  {Bi2Te3}: Topologically protected surface states and hexagonal warping
  effects},\ }\href {https://doi.org/10.1103/PhysRevLett.104.016401} {\bibfield
   {journal} {\bibinfo  {journal} {Phys. Rev. Lett.}\ }\textbf {\bibinfo
  {volume} {104}},\ \bibinfo {pages} {016401} (\bibinfo {year}
  {2010})}\BibitemShut {NoStop}%
\bibitem [{\citenamefont {Reis}\ \emph {et~al.}(2017)\citenamefont {Reis},
  \citenamefont {Li}, \citenamefont {Dudy}, \citenamefont {Bauernfeind},
  \citenamefont {Glass}, \citenamefont {Hanke}, \citenamefont {Thomale},
  \citenamefont {Schäfer},\ and\ \citenamefont {Claessen}}]{Reis2017}%
  \BibitemOpen
  \bibfield  {author} {\bibinfo {author} {\bibfnamefont {F.}~\bibnamefont
  {Reis}}, \bibinfo {author} {\bibfnamefont {G.}~\bibnamefont {Li}}, \bibinfo
  {author} {\bibfnamefont {L.}~\bibnamefont {Dudy}}, \bibinfo {author}
  {\bibfnamefont {M.}~\bibnamefont {Bauernfeind}}, \bibinfo {author}
  {\bibfnamefont {S.}~\bibnamefont {Glass}}, \bibinfo {author} {\bibfnamefont
  {W.}~\bibnamefont {Hanke}}, \bibinfo {author} {\bibfnamefont
  {R.}~\bibnamefont {Thomale}}, \bibinfo {author} {\bibfnamefont
  {J.}~\bibnamefont {Schäfer}},\ and\ \bibinfo {author} {\bibfnamefont
  {R.}~\bibnamefont {Claessen}},\ }\bibfield  {title} {\bibinfo {title}
  {Bismuthene on a {SiC} substrate: {A} candidate for a high-temperature
  quantum spin {Hall} material},\ }\href
  {https://doi.org/10.1126/science.aai8142} {\bibfield  {journal} {\bibinfo
  {journal} {Science}\ }\textbf {\bibinfo {volume} {357}},\ \bibinfo {pages}
  {287} (\bibinfo {year} {2017})}\BibitemShut {NoStop}%
\bibitem [{\citenamefont {Schneider}\ \emph {et~al.}(2021)\citenamefont
  {Schneider}, \citenamefont {Beck}, \citenamefont {Posske}, \citenamefont
  {Crawford}, \citenamefont {Mascot}, \citenamefont {Rachel}, \citenamefont
  {Wiesendanger},\ and\ \citenamefont {Wiebe}}]{Schneider2021}%
  \BibitemOpen
  \bibfield  {author} {\bibinfo {author} {\bibfnamefont {L.}~\bibnamefont
  {Schneider}}, \bibinfo {author} {\bibfnamefont {P.}~\bibnamefont {Beck}},
  \bibinfo {author} {\bibfnamefont {T.}~\bibnamefont {Posske}}, \bibinfo
  {author} {\bibfnamefont {D.}~\bibnamefont {Crawford}}, \bibinfo {author}
  {\bibfnamefont {E.}~\bibnamefont {Mascot}}, \bibinfo {author} {\bibfnamefont
  {S.}~\bibnamefont {Rachel}}, \bibinfo {author} {\bibfnamefont
  {R.}~\bibnamefont {Wiesendanger}},\ and\ \bibinfo {author} {\bibfnamefont
  {J.}~\bibnamefont {Wiebe}},\ }\bibfield  {title} {\bibinfo {title}
  {Topological {Shiba} bands in artificial spin chains on superconductors},\
  }\href {https://doi.org/10.1038/s41567-021-01234-y} {\bibfield  {journal}
  {\bibinfo  {journal} {Nat. Phys.}\ }\textbf {\bibinfo {volume} {17}},\
  \bibinfo {pages} {943} (\bibinfo {year} {2021})}\BibitemShut {NoStop}%
\bibitem [{\citenamefont {Bruus}\ and\ \citenamefont
  {Flensberg}(2004)}]{Bruus2004}%
  \BibitemOpen
  \bibfield  {author} {\bibinfo {author} {\bibfnamefont {H.}~\bibnamefont
  {Bruus}}\ and\ \bibinfo {author} {\bibfnamefont {K.}~\bibnamefont
  {Flensberg}},\ }\href {https://books.google.co.in/books?id=CktuBAAAQBAJ}
  {\emph {\bibinfo {title} {Many-Body Quantum Theory in Condensed Matter
  Physics: An Introduction}}},\ Oxford Graduate Texts\ (\bibinfo  {publisher}
  {OUP Oxford},\ \bibinfo {year} {2004})\BibitemShut {NoStop}%
\bibitem [{\citenamefont {Li}\ \emph {et~al.}(2014)\citenamefont {Li},
  \citenamefont {Xu},\ and\ \citenamefont {Chen}}]{Generalized}%
  \BibitemOpen
  \bibfield  {author} {\bibinfo {author} {\bibfnamefont {L.}~\bibnamefont
  {Li}}, \bibinfo {author} {\bibfnamefont {Z.}~\bibnamefont {Xu}},\ and\
  \bibinfo {author} {\bibfnamefont {S.}~\bibnamefont {Chen}},\ }\bibfield
  {title} {\bibinfo {title} {Topological phases of generalized
  {Su-Schrieffer-Heeger} models},\ }\href
  {https://doi.org/10.1103/PhysRevB.89.085111} {\bibfield  {journal} {\bibinfo
  {journal} {Phys. Rev. B}\ }\textbf {\bibinfo {volume} {89}},\ \bibinfo
  {pages} {085111} (\bibinfo {year} {2014})}\BibitemShut {NoStop}%
\bibitem [{\citenamefont {Qi}\ \emph {et~al.}(2006)\citenamefont {Qi},
  \citenamefont {Wu},\ and\ \citenamefont {Zhang}}]{QWZ2006}%
  \BibitemOpen
  \bibfield  {author} {\bibinfo {author} {\bibfnamefont {X.-L.}\ \bibnamefont
  {Qi}}, \bibinfo {author} {\bibfnamefont {Y.-S.}\ \bibnamefont {Wu}},\ and\
  \bibinfo {author} {\bibfnamefont {S.-C.}\ \bibnamefont {Zhang}},\ }\bibfield
  {title} {\bibinfo {title} {Topological quantization of the spin {Hall} effect
  in two-dimensional paramagnetic semiconductors},\ }\href
  {https://doi.org/10.1103/PhysRevB.74.085308} {\bibfield  {journal} {\bibinfo
  {journal} {Phys. Rev. B}\ }\textbf {\bibinfo {volume} {74}},\ \bibinfo
  {pages} {085308} (\bibinfo {year} {2006})}\BibitemShut {NoStop}%
\bibitem [{\citenamefont {Alicea}(2012)}]{AliceaReview2012}%
  \BibitemOpen
  \bibfield  {author} {\bibinfo {author} {\bibfnamefont {J.}~\bibnamefont
  {Alicea}},\ }\bibfield  {title} {\bibinfo {title} {New directions in the
  pursuit of {Majorana} fermions in solid state systems},\ }\href
  {http://stacks.iop.org/0034-4885/75/i=7/a=076501} {\bibfield  {journal}
  {\bibinfo  {journal} {Rep. Prog. Phys.}\ }\textbf {\bibinfo {volume} {75}},\
  \bibinfo {pages} {076501} (\bibinfo {year} {2012})}\BibitemShut {NoStop}%
\bibitem [{\citenamefont {Su}\ \emph {et~al.}(1979)\citenamefont {Su},
  \citenamefont {Schrieffer}, ,\ and\ \citenamefont {Heeger}}]{SSH}%
  \BibitemOpen
  \bibfield  {author} {\bibinfo {author} {\bibfnamefont {W.~P.}\ \bibnamefont
  {Su}}, \bibinfo {author} {\bibfnamefont {J.~R.}\ \bibnamefont {Schrieffer}},
  ,\ and\ \bibinfo {author} {\bibfnamefont {A.~J.}\ \bibnamefont {Heeger}},\
  }\bibfield  {title} {\bibinfo {title} {Solitons in polyacetylene},\ }\href
  {https://doi.org/10.1103/PhysRevLett.42.1698} {\bibfield  {journal} {\bibinfo
   {journal} {Phys. Rev. Lett.}\ }\textbf {\bibinfo {volume} {42}},\ \bibinfo
  {pages} {1698} (\bibinfo {year} {1979})}\BibitemShut {NoStop}%
\bibitem [{\citenamefont {Lee}\ \emph {et~al.}(2018)\citenamefont {Lee},
  \citenamefont {Imhof}, \citenamefont {Berger}, \citenamefont {Bayer},
  \citenamefont {Brehm}, \citenamefont {Molenkamp}, \citenamefont {Kiessling},\
  and\ \citenamefont {Thomale}}]{TopolectricalCircuits}%
  \BibitemOpen
  \bibfield  {author} {\bibinfo {author} {\bibfnamefont {C.~H.}\ \bibnamefont
  {Lee}}, \bibinfo {author} {\bibfnamefont {S.}~\bibnamefont {Imhof}}, \bibinfo
  {author} {\bibfnamefont {C.}~\bibnamefont {Berger}}, \bibinfo {author}
  {\bibfnamefont {F.}~\bibnamefont {Bayer}}, \bibinfo {author} {\bibfnamefont
  {J.}~\bibnamefont {Brehm}}, \bibinfo {author} {\bibfnamefont {L.~W.}\
  \bibnamefont {Molenkamp}}, \bibinfo {author} {\bibfnamefont {T.}~\bibnamefont
  {Kiessling}},\ and\ \bibinfo {author} {\bibfnamefont {R.}~\bibnamefont
  {Thomale}},\ }\bibfield  {title} {\bibinfo {title} {Topolectrical circuits},\
  }\href {https://doi.org/10.1038/s42005-018-0035-2} {\bibfield  {journal}
  {\bibinfo  {journal} {Commun. Phys.}\ }\textbf {\bibinfo {volume} {39}}
  (\bibinfo {year} {2018})}\BibitemShut {NoStop}%
\bibitem [{\citenamefont {Haldane}(1988)}]{Haldane}%
  \BibitemOpen
  \bibfield  {author} {\bibinfo {author} {\bibfnamefont {F.~D.~M.}\
  \bibnamefont {Haldane}},\ }\bibfield  {title} {\bibinfo {title} {Model for a
  quantum {Hall} effect with-out {Landau} levels: condensed-matter realization
  of the ‘parity anomaly’},\ }\href
  {https://doi.org/10.1103/PhysRevLett.61.2015} {\bibfield  {journal} {\bibinfo
   {journal} {Phys. Rev. Lett.}\ }\textbf {\bibinfo {volume} {61}},\ \bibinfo
  {pages} {085111} (\bibinfo {year} {1988})}\BibitemShut {NoStop}%
\bibitem [{\citenamefont {Damm}\ \emph {et~al.}(2017)\citenamefont {Damm},
  \citenamefont {Dung}, \citenamefont {Vewinger}, \citenamefont {Weitz},\ and\
  \citenamefont {Schmitt}}]{Damm2017}%
  \BibitemOpen
  \bibfield  {author} {\bibinfo {author} {\bibfnamefont {T.}~\bibnamefont
  {Damm}}, \bibinfo {author} {\bibfnamefont {D.}~\bibnamefont {Dung}}, \bibinfo
  {author} {\bibfnamefont {F.}~\bibnamefont {Vewinger}}, \bibinfo {author}
  {\bibfnamefont {M.}~\bibnamefont {Weitz}},\ and\ \bibinfo {author}
  {\bibfnamefont {J.}~\bibnamefont {Schmitt}},\ }\bibfield  {title} {\bibinfo
  {title} {First-order spatial coherence measurements in a thermalized
  two-dimensional photonic quantum gas},\ }\href
  {https://doi.org/10.1038/s41467-017-00270-8} {\bibfield  {journal} {\bibinfo
  {journal} {Nat. Commun.}\ }\textbf {\bibinfo {volume} {8}},\ \bibinfo {pages}
  {158} (\bibinfo {year} {2017})}\BibitemShut {NoStop}%
\bibitem [{\citenamefont {Ligthart}\ \emph {et~al.}(2024)\citenamefont
  {Ligthart}, \citenamefont {Herrera}, \citenamefont {Visser}, \citenamefont
  {Vlasblom}, \citenamefont {Bercioux},\ and\ \citenamefont
  {Swart}}]{Ligthart2024}%
  \BibitemOpen
  \bibfield  {author} {\bibinfo {author} {\bibfnamefont {R.~A.~M.}\
  \bibnamefont {Ligthart}}, \bibinfo {author} {\bibfnamefont {M.~A.~J.}\
  \bibnamefont {Herrera}}, \bibinfo {author} {\bibfnamefont {A.}~\bibnamefont
  {Visser}}, \bibinfo {author} {\bibfnamefont {A.}~\bibnamefont {Vlasblom}},
  \bibinfo {author} {\bibfnamefont {D.}~\bibnamefont {Bercioux}},\ and\
  \bibinfo {author} {\bibfnamefont {I.}~\bibnamefont {Swart}},\ }\bibfield
  {title} {\bibinfo {title} {Experimental determination of {Wannier} centers in
  {1D} topological chains},\ }\href {https://arxiv.org/abs/2407.14465} {\
  (\bibinfo {year} {2024})},\ \Eprint {https://arxiv.org/abs/2407.14465}
  {arXiv:2407.14465 [cond-mat.mes-hall]} \BibitemShut {NoStop}%
\bibitem [{\citenamefont {Hsieh}\ \emph {et~al.}(2009)\citenamefont {Hsieh},
  \citenamefont {Xia}, \citenamefont {Wray}, \citenamefont {Qian},
  \citenamefont {Pal}, \citenamefont {Dil}, \citenamefont {Osterwalder},
  \citenamefont {Meier}, \citenamefont {Bihlmayer}, \citenamefont {Kane},
  \citenamefont {Hor}, \citenamefont {Cava},\ and\ \citenamefont
  {Hasan}}]{Hsieh2009}%
  \BibitemOpen
  \bibfield  {author} {\bibinfo {author} {\bibfnamefont {D.}~\bibnamefont
  {Hsieh}}, \bibinfo {author} {\bibfnamefont {Y.}~\bibnamefont {Xia}}, \bibinfo
  {author} {\bibfnamefont {L.}~\bibnamefont {Wray}}, \bibinfo {author}
  {\bibfnamefont {D.}~\bibnamefont {Qian}}, \bibinfo {author} {\bibfnamefont
  {A.}~\bibnamefont {Pal}}, \bibinfo {author} {\bibfnamefont {J.~H.}\
  \bibnamefont {Dil}}, \bibinfo {author} {\bibfnamefont {J.}~\bibnamefont
  {Osterwalder}}, \bibinfo {author} {\bibfnamefont {F.}~\bibnamefont {Meier}},
  \bibinfo {author} {\bibfnamefont {G.}~\bibnamefont {Bihlmayer}}, \bibinfo
  {author} {\bibfnamefont {C.~L.}\ \bibnamefont {Kane}}, \bibinfo {author}
  {\bibfnamefont {Y.~S.}\ \bibnamefont {Hor}}, \bibinfo {author} {\bibfnamefont
  {R.~J.}\ \bibnamefont {Cava}},\ and\ \bibinfo {author} {\bibfnamefont
  {M.~Z.}\ \bibnamefont {Hasan}},\ }\bibfield  {title} {\bibinfo {title}
  {Observation of unconventional quantum spin textures in topological
  insulators},\ }\href {https://doi.org/10.1126/science.1167733} {\bibfield
  {journal} {\bibinfo  {journal} {Science}\ }\textbf {\bibinfo {volume}
  {323}},\ \bibinfo {pages} {919} (\bibinfo {year} {2009})}\BibitemShut
  {NoStop}%
\bibitem [{\citenamefont {Jozwiak}\ \emph {et~al.}(2011)\citenamefont
  {Jozwiak}, \citenamefont {Chen}, \citenamefont {Fedorov}, \citenamefont
  {Analytis}, \citenamefont {Rotundu}, \citenamefont {Schmid}, \citenamefont
  {Denlinger}, \citenamefont {Chuang}, \citenamefont {Lee}, \citenamefont
  {Fisher}, \citenamefont {Birgeneau}, \citenamefont {Shen}, \citenamefont
  {Hussain},\ and\ \citenamefont {Lanzara}}]{Jozwiak2011}%
  \BibitemOpen
  \bibfield  {author} {\bibinfo {author} {\bibfnamefont {C.}~\bibnamefont
  {Jozwiak}}, \bibinfo {author} {\bibfnamefont {Y.~L.}\ \bibnamefont {Chen}},
  \bibinfo {author} {\bibfnamefont {A.~V.}\ \bibnamefont {Fedorov}}, \bibinfo
  {author} {\bibfnamefont {J.~G.}\ \bibnamefont {Analytis}}, \bibinfo {author}
  {\bibfnamefont {C.~R.}\ \bibnamefont {Rotundu}}, \bibinfo {author}
  {\bibfnamefont {A.~K.}\ \bibnamefont {Schmid}}, \bibinfo {author}
  {\bibfnamefont {J.~D.}\ \bibnamefont {Denlinger}}, \bibinfo {author}
  {\bibfnamefont {Y.-D.}\ \bibnamefont {Chuang}}, \bibinfo {author}
  {\bibfnamefont {D.-H.}\ \bibnamefont {Lee}}, \bibinfo {author} {\bibfnamefont
  {I.~R.}\ \bibnamefont {Fisher}}, \bibinfo {author} {\bibfnamefont {R.~J.}\
  \bibnamefont {Birgeneau}}, \bibinfo {author} {\bibfnamefont {Z.-X.}\
  \bibnamefont {Shen}}, \bibinfo {author} {\bibfnamefont {Z.}~\bibnamefont
  {Hussain}},\ and\ \bibinfo {author} {\bibfnamefont {A.}~\bibnamefont
  {Lanzara}},\ }\bibfield  {title} {\bibinfo {title} {Widespread spin
  polarization effects in photoemission from topological insulators},\ }\href
  {https://doi.org/10.1103/PhysRevB.84.165113} {\bibfield  {journal} {\bibinfo
  {journal} {Phys. Rev. B}\ }\textbf {\bibinfo {volume} {84}},\ \bibinfo
  {pages} {165113} (\bibinfo {year} {2011})}\BibitemShut {NoStop}%
\bibitem [{\citenamefont {Zhu}\ \emph {et~al.}(2013)\citenamefont {Zhu},
  \citenamefont {Veenstra}, \citenamefont {Levy}, \citenamefont {Ubaldini},
  \citenamefont {Syers}, \citenamefont {Butch}, \citenamefont {Paglione},
  \citenamefont {Haverkort}, \citenamefont {Elfimov},\ and\ \citenamefont
  {Damascelli}}]{Zhu2013}%
  \BibitemOpen
  \bibfield  {author} {\bibinfo {author} {\bibfnamefont {Z.-H.}\ \bibnamefont
  {Zhu}}, \bibinfo {author} {\bibfnamefont {C.~N.}\ \bibnamefont {Veenstra}},
  \bibinfo {author} {\bibfnamefont {G.}~\bibnamefont {Levy}}, \bibinfo {author}
  {\bibfnamefont {A.}~\bibnamefont {Ubaldini}}, \bibinfo {author}
  {\bibfnamefont {P.}~\bibnamefont {Syers}}, \bibinfo {author} {\bibfnamefont
  {N.~P.}\ \bibnamefont {Butch}}, \bibinfo {author} {\bibfnamefont
  {J.}~\bibnamefont {Paglione}}, \bibinfo {author} {\bibfnamefont {M.~W.}\
  \bibnamefont {Haverkort}}, \bibinfo {author} {\bibfnamefont {I.~S.}\
  \bibnamefont {Elfimov}},\ and\ \bibinfo {author} {\bibfnamefont
  {A.}~\bibnamefont {Damascelli}},\ }\bibfield  {title} {\bibinfo {title}
  {Layer-by-layer entangled spin-orbital texture of the topological surface
  state in {${\mathrm{Bi}}_{2}{\mathrm{Se}}_{3}$}},\ }\href
  {https://doi.org/10.1103/PhysRevLett.110.216401} {\bibfield  {journal}
  {\bibinfo  {journal} {Phys. Rev. Lett.}\ }\textbf {\bibinfo {volume} {110}},\
  \bibinfo {pages} {216401} (\bibinfo {year} {2013})}\BibitemShut {NoStop}%
\bibitem [{\citenamefont {Jozwiak}\ \emph {et~al.}(2016)\citenamefont
  {Jozwiak}, \citenamefont {Sobota}, \citenamefont {Gotlieb}, \citenamefont
  {Kemper}, \citenamefont {Rotundu}, \citenamefont {Birgeneau}, \citenamefont
  {Hussain}, \citenamefont {Lee}, \citenamefont {Shen},\ and\ \citenamefont
  {Lanzara}}]{Jozwiak2016}%
  \BibitemOpen
  \bibfield  {author} {\bibinfo {author} {\bibfnamefont {C.}~\bibnamefont
  {Jozwiak}}, \bibinfo {author} {\bibfnamefont {J.}~\bibnamefont {Sobota}},
  \bibinfo {author} {\bibfnamefont {K.}~\bibnamefont {Gotlieb}}, \bibinfo
  {author} {\bibfnamefont {A.~F.}\ \bibnamefont {Kemper}}, \bibinfo {author}
  {\bibfnamefont {C.~R.}\ \bibnamefont {Rotundu}}, \bibinfo {author}
  {\bibfnamefont {R.~J.}\ \bibnamefont {Birgeneau}}, \bibinfo {author}
  {\bibfnamefont {Z.}~\bibnamefont {Hussain}}, \bibinfo {author} {\bibfnamefont
  {D.-H.}\ \bibnamefont {Lee}}, \bibinfo {author} {\bibfnamefont {Z.-X.}\
  \bibnamefont {Shen}},\ and\ \bibinfo {author} {\bibfnamefont
  {A.}~\bibnamefont {Lanzara}},\ }\bibfield  {title} {\bibinfo {title}
  {Spin-polarized surface resonances accompanying topological surface state
  formation},\ }\href {https://doi.org/10.1038/ncomms13143} {\bibfield
  {journal} {\bibinfo  {journal} {Nat. Commun.}\ }\textbf {\bibinfo {volume}
  {7}},\ \bibinfo {pages} {13143} (\bibinfo {year} {2016})}\BibitemShut
  {NoStop}%
\bibitem [{\citenamefont {Bansil}\ and\ \citenamefont
  {Lindroos}(1999)}]{Bansil1999}%
  \BibitemOpen
  \bibfield  {author} {\bibinfo {author} {\bibfnamefont {A.}~\bibnamefont
  {Bansil}}\ and\ \bibinfo {author} {\bibfnamefont {M.}~\bibnamefont
  {Lindroos}},\ }\bibfield  {title} {\bibinfo {title} {Importance of matrix
  elements in the {ARPES} spectra of {BISCO}},\ }\href
  {https://doi.org/10.1103/PhysRevLett.83.5154} {\bibfield  {journal} {\bibinfo
   {journal} {Phys. Rev. Lett.}\ }\textbf {\bibinfo {volume} {83}},\ \bibinfo
  {pages} {5154} (\bibinfo {year} {1999})}\BibitemShut {NoStop}%
\bibitem [{\citenamefont {Vyas}\ and\ \citenamefont {Roy}(2021)}]{Vyas21}%
  \BibitemOpen
  \bibfield  {author} {\bibinfo {author} {\bibfnamefont {V.~M.}\ \bibnamefont
  {Vyas}}\ and\ \bibinfo {author} {\bibfnamefont {D.}~\bibnamefont {Roy}},\
  }\bibfield  {title} {\bibinfo {title} {Topological aspects of periodically
  driven non-{Hermitian Su-Schrieffer-Heeger} model},\ }\href
  {https://doi.org/10.1103/PhysRevB.103.075441} {\bibfield  {journal} {\bibinfo
   {journal} {Phys. Rev. B}\ }\textbf {\bibinfo {volume} {103}},\ \bibinfo
  {pages} {075441} (\bibinfo {year} {2021})}\BibitemShut {NoStop}%
\bibitem [{\citenamefont {Nehra}\ and\ \citenamefont {Roy}(2022)}]{ritu}%
  \BibitemOpen
  \bibfield  {author} {\bibinfo {author} {\bibfnamefont {R.}~\bibnamefont
  {Nehra}}\ and\ \bibinfo {author} {\bibfnamefont {D.}~\bibnamefont {Roy}},\
  }\bibfield  {title} {\bibinfo {title} {Topology of multipartite
  non-{Hermitian} one-dimensional systems},\ }\href
  {https://doi.org/10.1103/PhysRevB.105.195407} {\bibfield  {journal} {\bibinfo
   {journal} {Phys. Rev. B}\ }\textbf {\bibinfo {volume} {105}},\ \bibinfo
  {pages} {195407} (\bibinfo {year} {2022})}\BibitemShut {NoStop}%
\bibitem [{\citenamefont {Bernevig}\ and\ \citenamefont
  {Hughes}(2013)}]{Bernevig}%
  \BibitemOpen
  \bibfield  {author} {\bibinfo {author} {\bibfnamefont {B.~A.}\ \bibnamefont
  {Bernevig}}\ and\ \bibinfo {author} {\bibfnamefont {T.~L.}\ \bibnamefont
  {Hughes}},\ }\href@noop {} {\emph {\bibinfo {title} {Topological Insulators
  and Topological Superconductors}}}\ (\bibinfo  {publisher} {Princeton
  University Press},\ \bibinfo {year} {2013})\BibitemShut {NoStop}%
\end{thebibliography}%

\end{document}